\documentclass{article}

\usepackage{arxiv}

\usepackage[utf8]{inputenc} 
\usepackage[T1]{fontenc}    
\usepackage{url}            
\usepackage{booktabs}       
\usepackage{amsfonts}       
\usepackage{nicefrac}       
\usepackage{lipsum}		
\usepackage{graphicx}
\usepackage{doi}

\usepackage{amsthm,amsmath} 
\usepackage{mathrsfs} 
\usepackage{amsmath}
\usepackage{amssymb}
\usepackage{latexsym}
\usepackage{bm}
\usepackage{color}
\usepackage{url} 
\newcommand{\choosefont}[1]{\fontfamily{#1}\selectfont}

\usepackage{mathtools} 
\usepackage{algorithmic} 
\usepackage{algorithm}
\usepackage{setspace}

\theoremstyle{plain}

\newtheorem*{thm*}{Theorem}

\theoremstyle{definition}

\newcommand{\argmin}{\mathop{\rm arg~min}\limits}

\title{Causal rule ensemble method for estimating heterogeneous treatment effect with consideration of main effects}

\author{ \href{https://orcid.org/0000-0001-9944-9569}{\includegraphics[scale=0.06]{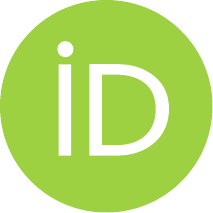}\hspace{1mm}Mayu Hiraishi} \\
	Clinical Study Support Center\\
	Wakayama Medical University Hospital\\
    Graduate School of Culture and Information Science\\
    Doshisha University\\
	\And
	\href{https://orcid.org/0000-0002-1563-7181}{\includegraphics[scale=0.06]{orcid.pdf}\hspace{1mm}Ke Wan} \\
	Department of Medical Data Science\\
	Wakayama Medical University\\
    \And
	\href{https://orcid.org/0000-0001-9621-5871}{\includegraphics[scale=0.06]{orcid.pdf}\hspace{1mm}Kensuke Tanioka} \\
	Department of Biomedical Sciences and Informatics\\
	Doshisha University\\
    \And
	\href{https://orcid.org/0000-0002-1408-2655}{\includegraphics[scale=0.06]{orcid.pdf}\hspace{1mm}Hiroshi Yadohisa} \\
	Department of Culture and Information Science\\
	Doshisha University\\
    \And
	\href{https://orcid.org/0000-0002-2256-0282}{\includegraphics[scale=0.06]{orcid.pdf}\hspace{1mm}Toshio Shimokawa} \\
	Department of Medical Data Science\\
	Wakayama Medical University\\
}

\renewcommand{\shorttitle}{\textit{arXiv} Template}

\begin{document}
\maketitle

\begin{abstract}
This study proposes a novel framework based on the RuleFit method to estimate Heterogeneous Treatment Effect (HTE) in a randomized clinical trial. 
To achieve this, we adopted S-learner of the metaalgorithm for our proposed framework. 
The proposed method incorporates a rule term for the main effect and treatment effect, which allows HTE to be interpretable form of rule. 
By including a main effect term in the proposed model, the selected rule is represented as an HTE that excludes other effects.
We confirmed a performance equivalent to that of another ensemble learning methods through numerical simulation and demonstrated the interpretation of the proposed method from a real data application. 
\end{abstract}


\section{Introduction}
\label{intro}
Randomized controlled clinical trials are conducted to verify the effect of new treatments and interventions as compared to standard  treatments. The average treatment effect is commonly used to evaluate the difference between the outcomes of new treatments and existing ones [Holland, 1986, Gail and Simon, 1985]. However, the treatment effect is not always homogeneous to the overall population, and varies to some individual characteristics. 
Heterogeneous treatment effect (HTE) has received wide attention in recent years.
HTE mainly focuses on the expected change in treatment outcomes by estimating the treatment effect at the individual level and capturing the specific characteristics of subgroups that receive effective new treatments. 
Various machine learning methodologies have been developed for for the estimating the HTE. Tree-based methods are well-suited for handling large-scale data and enabling flexible modeling with various levels of covariates' measurements, compared to conventional statistical methods [Wager and Athey, 2018]. 
For example, regression tree construction based on CART [Breiman et al., 1984] was proposed in Su et al. [2009], Athey and Imbens [2016]. In a forest-based algorithm, Random Forest [Breiman, 2001] is extended to causal effect estimation   
[Wager and Athey, 2018, Athey et al., 2019].
Powers et al. [2018] proposed three methods in a framework of the conditional outcome difference. In Bayesian approaches, 
Bayesian additive regression trees (BART) [Chipman et al., 2010]-based methods have been developled for HTE estimation [Hill, 2011, Hahn et al., 2020]. 
Although ensemble learning models have demonstrated significant predictive capabilities, the interpretation of variable contributions to the predicted values lacks clarity, which indicates that the model is a black box.

To address this challenge, model-interpretable methods have been proposed. The RuleFit method introduced by Friedman and Popescu [2008] ues a nonparametric tree-based ensemble technique that can be expressed as a linear combination of base functions. 
This method generates functions based on "rules" from the paths of each root to a terminal node in each decision tree, and these rules can be extracted as the style of the rule, i.e., ''weight $>
64.8$ kg \& height $ \leq 173.6$ cm''.
These rules help determining the relationship between the characteristics of subgroups and the effect of the treatment. 
The RuleFit method has been applied to the framework of HTE in several studies. 
Bargagli-Stoffi et al. [2023] proposed the causal rule ensemble (CRE), which uses the RuleFit to extract the interpretable HTE as the form of rule after estimating HTE by another ensemble method. However, this method does not use RuleFit for the estimation of HTE itself.

Thus, this study proposes a novel framework based on the RuleFit to estimate the interpretable HTE. 
The proposed framework assumes that the estimated HTE can be expressed as a linear combination of coefficients and rules to interpret HTE between target treatment group and control group. 
Then, we interpret the characteristics represented by the obtained rules in terms of how they affect HTE. 
The proposed method includes a main effect term in addition to treatment effect terms to express HTE as a linear combination. 
If the main effect term is not in the model, the estimated value may contain both the main effect and HTE, making it difficult to accurately evaluate the specific effects of the treatment. 
Therefore, including a main effect term in the model allows HTE to be interpreted in the form of rules.  

To realize this model, the proposed method employs S-learner in Metaalgorithm [Künzel et al., 2019].
Metaalgorithm, also called the meta-learner, is a framework in causal inference that estimates HTE in the machine learning literature.  
While other learner methods such as T-learner (where "T" denotes two) are formed by two models by each treatment group, S-learner, where "S" denotes "single", can provide a single model for estimating HTE. 
The proposed method is based on the S-learner due to the structure of the framework, which uses rule function as the base function to estimate HTE. 
In the S-learner and T-learner, HTE is estimated directly using the predictions of the regression model fitted to the responses.
However, becaused the T-learner constructs the model separately between each treatment group, it is difficult to consider, in the process of estimating HTE, the cases wherein the treatment and control group share common effects [Künzel et al., 2019]. 
The results of the numerical simulations of Nie and Wager [2020] show that the S-learner had better performance than the T-learner in some situations. 
Moreover, when calculating the HTE based on the  difference between each treatment group in the T-learner, the common effects between the two treatment groups were estimated separately as well. 
Therefore, interpreting the treatment-specific effects is difficult because the estimated treatment effect includes the main effect.
S-learner is the only learner that allows the construction of a model with main effect and interaction terms. 
Additionally, in the proposed method, the rule term and its coefficients for the main effect 
are common between treatment group and control group.
The S-learner can construct this structure, consider the main effects, and interpret the HTE in the form of rules. Therefore, the S-learner is appropriate for the proposed method. 

Moreover, to obtain an interpretable HTE, i.e., to represent the estimated HTE as a linear combination, we incorporate the idea of shared basis proposed by Powers et al. [2018] into the proposed framework. 
Shared-basis shares the base function of conditional mean regression between the model of two treatment groups to compare HTE without excluding selection bias of base function. 
This concept has been discussed within the framework of the T-learner. The proposed framework adopts the idea of a shared basis into the S-learner framework to ensure the comparability of HTE between the two treatment groups. 
In particular, by sharing the same rules related to HTE between the target treatment group and control group, the calculation results of HTE can be described as a linear combination of coefficients and rules.  
To accomplish this, we use group lasso [Yuan and Lin, 2006] instead of the lasso [Tibshirani, 1996] used in the conventional RuleFit method. 
In the metaalgorithm framework, S-learner is appropriate for building models that adopt the idea of shared basis. We incorporate this advantage of the S-learner in the proposed method to easily select the same rules considering it contains the rule terms of the two treatment groups in one model. 
Wan et al. [2023] proposed a RuleFit-based method to estimate HTE; however, it does not consider the main effect. In contrast, the proposed method considers the main effect in estimating HTE, thereby allowing for a more refined interpretation of the treatment effect. 

In Section $2$, we explain HTE and conventional RuleFit in relation to the proposed method. Next, we introduce the framework and algorithm of the proposed method in Section $3$.
Then, we demonstrate the efficiency of the proposed method through numerical simulation in Section $4$. In Section $5$, we describe the application of the proposed method to real genetic data related to breast cancer. Based on the results, Section $6$ concludes the article.

\section{Related works}
We extend our proposed method using the RuleFit method [Friedman and Popescu, 2008] to estimate the heterogeneous treatment effect (HTE) in a randomized clinical trial. 
Before presenting our method, we explain HTE and RuleFit. \\

\subsection{Heterogeneous treatment effect (HTE)}
In randomized controlled clinical trials, the target treatments are compared with the standard treatments to test the effectiveness of new treatments. The average treatment effect (ATE) is typically used for estimation.
However, the ATE cannot detect the subgroups for whom the new treatment is more effective than the standard treatment owing to the average of the population. To identify subgroups, the heterogeneous treatment effect (HTE) focuses on the variability in treatment effects that may be attributed to patient factors [Gail and Simon, 1985].
Let $Y_i \ (i=1,2,\cdots, n)$ be the outcome variable, where $n$ is the number of subjects, $X_i=(X_{i1},X_{i2}, \cdots, X_{ip})^T \ (i=1, 2, \cdots, n)$ be random covariate vectors, where $p$ is the number of variables and $\cdot^T$ denotes the transpose, and $Z_i \in \{0,1\}$ be the allocation group in two levels, where $Z_i=1$ and $Z_i=0$ are the target treatment group and control treatment group, respectively.

Herein, we describe the settings used in this study. 
Each subject exhibited the only one response to treatment. \\
The treatment effect considering heterogeneity is defined as 

\begin{align}
\tau(\bm{x}_i) = \mu_1(\bm{x}_i) - \mu_0(\bm{x}_i)
\label{taut}
\end{align}
where
\begin{align}
\mu_1(\bm{x}_i) = \text{E}(Y_i|X_i=\bm{x}_i, Z_i=1), \ \mu_0(\bm{x}_i) = \text{E}(Y_i|X_i=\bm{x}_i, Z_i=0).
\label{hte1}
\end{align}
Here, $\bm{x}_i \in \mathbb{R}^p$ is the observed covariate vector. HTE is the difference in the conditional mean functions between two treatment groups. 
$\mu_1(\bm{x}_i)$ and $\mu_0(\bm{x}_i)$ denotes the expected respected response when subject $i$ assigned to the target treatment group and standard treatment group, respectively.
In this study, Eq. (1) and Eq. (2) are used to estimate the HTE.

\subsection{RuleFit}
\label{2.2}
RuleFit is a rule-based ensemble method proposed in Friedman and Popescu [2008]. 
This method can handle cases where the relationships between the outcomes and covariates are even nonlinear.
Furthermore, it ensures the interpretability of the results by extracting the generated rules.
Given the covariates $\bm{x}_i=(x_{i1},x_{i2}, \cdots, x_{ip})^T \ (i=1, 2, \cdots, n)$, the model of the RuleFit is defined as 
\begin{align}
F_{RF}(\bm{x}_i) = \beta_0 + \sum_{k=1}^K\beta_kr_k(\bm{x}_i)+ \sum_{j=1}^p \alpha_jl_j(\bm{x}_i) \ (i=1, 2, \cdots, n)
\label{rfmodel}
\end{align}
where $\beta_0 \in \mathbb{R}$ denotes the intercept, $\beta_k \in \mathbb{R} \ (k=1,2,\cdots, K)$ denotes the coefficients of the rule terms, and $\alpha_j \in \mathbb{R} \ (j=1,2,\cdots, p)$ denotes the coefficient of the linear terms. In Eq. (3), the RuleFit method a rule term and a linear term.
The rule term of $k$th rule is defined as the following function:
\begin{align*}
r_k (\bm{x}_i)= \prod_{j=1}^pI(x_{ij} \in  S_{jk})
\end{align*}
where $S_j$ is the set of all possible values of the covariates $x_{ij} (x_{ij} \in S_j )$ and 
$S_{jk} \subset S_j$, and
$I(\cdot)$ is an indicator function that returns $1$ if $x_{ij} \in S_{jk}$ is true; else it returns $0$. $S_{jk}$ can be defined from the interval $ ( x_{jk}^-,x_{jk}^+ ] \ (x_{jk}^-,x_{jk}^+ \in \mathbb{R})$ when $\bm{x}_j$ is an ordinal or scale variable. Rule- or tree-based ensembles have difficulty approximating linear structures, particularly when the number of training samples is insufficient. As a result, it may not generate sufficient rules to estimate the appropriate model [Friedman and Popescu, 2008]. To improve the accuracy and the interpretability, the RuleFit model adds a linear term as an additional basis functions based on the variable $j$.  To reduce the influence of the outliers of the covariates, the linear function $l_j(\cdot)$ is substituted for the "Winsorized" version to provide robustness. The "Winsorized" version of the linear function is defined as: 
\begin{align}
l_j(\bm{x}_i) = \min\bigl(\delta^+_j,\max(\delta_j^-,x_{ij})\bigr)
\label{rfl}
\end{align}
where $\delta^-_j$ and $\delta_j^+$ are the thresholds of the outliers, which are the $q \in (0, 1)$ and $(1-q)$ quantiles of variable $j$. Friedman and Popescu [2008] recommends $q \simeq 0.025$, which has been adopted it in this paper.
The coefficient vector of the linear term also depends on the scale. Therefore, Eq. (4) is normalized as:
\begin{align*}
l_j(\bm{x}_i) \leftarrow 0.4 \cdot \frac{l_j(\bm{x}_i)}{std(l_j(\bm{x}_i))} 
\end{align*}
where $std(l_j(\bm{x}_i))$ is the standard deviation of $l_j(\bm{x}_i)$. Here, $0.4$ is the average standard deviation of the rule under certain conditions [Friedman and Popescu, 2008, Fokkema, 2020].

\section{Proposed method}
In this section, we present the framework of the proposed method and the calculation of HTE based on the proposed method. Then, we explain the algorithm. 

\subsection{Framework of the proposed method} \label{sec3.1}
We define a model of the proposed method. We then explain the four steps for estimating the HTE based on the proposed method.

Let $y_i \in \mathbb{R}$ be the continuous outcome variable, $\bm{x}_i \in \mathbb{R}^p$ be the covariates,  $z_i \in \{0,1\}$ be the treatment group, $r^\dagger_{k^\dagger}(\cdot) \ (k^\dagger = 1, 2, \cdots, K^\dagger)$ be rule function for the main effect, $r^*_{k^*}(\cdot) \ (k^*=1,2,\cdots, K^*)$ be rule function for the treatment effect, and $l_j(\cdot) \ (j = 1, 2, \cdots, p)$ be the linear function.
Given $y_i, \bm{x}_i$, and $z_i$,
the model of the proposed method is defined as
\begin{align}
F(\bm{x}_i,z_i) = & 
\beta_0 + \sum_{k^\dagger=1}^{K^\dagger} \beta_{k^\dagger}r^\dagger_{k^\dagger}(\bm{x}_i) + \sum_{j=1}^p \alpha_j l_j(\bm{x}_i) 
+ \sum_{k^*=1}^{K^*}\beta_{k^*}^{(a)}r^*_{k^*}(\bm{x}_i) \cdot I (z_i=1)
+ \sum_{k^*=1}^{K^*}\beta_{k^*}^{(c)}r^*_{k^*}(\bm{x}_i) \cdot I (z_i=0),
\label{prp}
\end{align}
where 
$\beta_0$ is the intercept, $\beta_{k^\dagger} \ (k^\dagger=1, 2, \cdots, K^\dagger)$ is the coefficients of the rule term of the main effect, and $ \alpha_j \ (j=1, 2, \cdots, p)$ is the coefficients of the linear term of main effect. 

$\beta_{k^*}^{(a)} \ (k^*=1, 2, \cdots, K^*)$ is the coefficients of the rule term of the treatment effect in the target treatment group $z=1$, whereras $\beta_{k^*}^{(c)} \ (k^*=1, 2, \cdots, K^*)$ as that in the control treatment group $z=0$.
$I(\cdot)$ denotes the indication function.
The first, second, and third terms in Eq. (5) are associated with the main effect, which does not depend on the treatment.
In contrast, the fourth and fifth terms in Eq. (5) are those related to treatment.

To calculate the HTE, the proposed method follows the four steps below:\\
\noindent
{\bf STEP 1 :} 
Generation of a base function.

Given $\bigl(y_i, (\bm{x}_i, z_i)\bigr) \ (i=1, 2, \cdots, n)$ where $y_i$ is a continuous response variable, the $k$th base function explaining $y_i$ is generated as: 
\begin{align}
r_k (\bm{x}_i,z_i)= I(z_i \in V_{zk})\prod_{j=1}^pI(x_{ij} \in  S_{jk})
\label{prop1}
\end{align}
where $V_{zk} \subset\{1,0\}$ is a subset of all the possible values of the allocation groups. Eq. (6) can be generated following the procedure of the original RuleFit. 
Details of this step are provided in 3.2.1.

\noindent
{\bf STEP 2 :} 
Rule Sorting.

The set of all rule functions generated in ${\bf STEP 1}$ is denoted as $Rule$. 
In this step, all generated $K$ rules are divided into $K^\dagger$ rules associated with the main effect and $K^*$ rules associated with the treatments as 
\begin{align}
Rule^\dagger &= \{ r_k(\cdot,\cdot) | V_{zk} = \{1, 0\}\},  \\
Rule^* &= \{ r_k(\cdot,\cdot) | V_{zk} = \{ 1 \}  \vee V_{zk} =\{0\}\} \ (k=1,2,\cdots, K)
\label{prop2}
\end{align}
where $\#(Rule^\dagger) = K^\dagger$ and $\#(Rule^*) = K^*$. $\#$ denotes the cardinality of the set. 
$Rule^\dagger$ is the set of base functions related to the main effect and is denoted by $r^\dagger_{k^\dagger}(\bm{x}_i) =  r_k(\bm{x}_i,z_i) \in Rule^\dagger$. Additionally, $Rule^*$ is a set of base functions related to the treatment effects.
We set $r_{k^*}^*(\bm{x}_i) = \prod_{j=1}^p I(x_{ij} \in S_{jk^*})$, and $r^*_{k^*}(\bm{x}_i) \cdot I(z_i \in V_{zk^*})= r_k (\bm{x}_i, z_i)  \in Rule^*$ is defined.
Here, $Rule^\dagger \cup Rule^* = Rule$ and $Rule^\dagger \cap Rule^* = \phi$. 
The details of this step are as follows.

\noindent
{\bf STEP 3 :} 
Estimation of regression coefficients.  

Based on the base functions in ${\bf STEP 2}$, Eq. (5) is constructed. 
The optimization problem of the proposed method for {\bf STEP 2} is defined as:
\begin{align}
&L\bigl(\beta_0,\{\beta_{k^\dagger}\}_{k^\dagger=1}^{K^\dagger},\{\alpha_j\}_{j=1}^p, \{\bm{\beta}_{k^*}^{\ddagger}\}_{k^*=1}^{K^*}\bigr) \nonumber\\
&= \frac{1}{2}\sum_{i=1}^n (y_i - F(\bm{x}_i, z_i))^2 
+ \lambda \left(\sum_{k^\dagger=1}^{K^\dagger} | \beta_{k^\dagger}| + \sum_{j=1}^p|\alpha_j |
+ \sqrt{2}\sum_{k^*=1}^{K^*} \| \bm{\beta}^{\ddagger}_{k^*} \|_2 \right)
\rightarrow \text{min}
\label{prpf}
\end{align}
where $\|\cdot \|_2$ is the L$2$ norm.
$\beta_{k^\dagger} \ (k^\dagger=1,2, \cdots, K^\dagger)$ and $\alpha_j \ (j=1,2, \cdots, p)$ are the coefficient vectors of the rule term and the linear term effects, respectively for the main effect. $\bm{\beta}_{k^*}^{\ddagger} = (\beta_{k^*}^{(a)}, \beta_{k^*}^{(c)})^T \ (k^*=1,2,\cdots, K^*)$ is the set of coefficient vectors related to the treatment. $\beta_{k^*}^{(a)}$ and $\beta_{k^*}^{(c)}$ are the parameter of the target treatment group and the control group, respectively.\\ 
The conventional RuleFit method uses lasso [Tibshirani, 1996] to prune the base learners, whereas the proposed method uses group lasso [Yuan and Lin, 2006] to prune the rule terms for HTE. This allows selection of the same rules between the base functions in the two treatment groups and is expected to ensure comparability of the two treatment groups. The details of this step are presented in Section 3.2.2.

\noindent
{\bf STEP 4 :} 
Calculation of HTE. 

Using the parameters estimated from {\bf STEP 3}, the HTE is computed based on the model and allocation groups. 
The details are presented in Section 3.2.3.

\subsection{Algorithm} 
In this subsection, we explain rule generation, rule sorting, estimation of the regression coefficients, and calculation of the HTE using the algorithm of the proposed method introduced in Section 3.1. 

\subsubsection{Rule generation related to both main effect and treatment effect }
In 3.2.1, the base function $\{r_k(\cdot)\}_{k=1}^K$ is generated from the covariate $\bm{x}$ and allocation group $z$ of the training data $\{y_i, (\bm{x}_i, z_i)\}_{i=1}^{n_1}$, where $n_1$ is the number of subject in the training data.  The details are presented in Algorithm $1$. 
Model $H$ is formed as $H(\bm{x}_i, z_i) = \{h_m(\bm{x}_i, z_i)\}_{m=1}^M (i=1,2, \cdots, n_1; \ m=1,2,\cdots, M)$, where $M$ is the number of tree-based learner $h_m$.
$M$, $\bar{L} \ (\bar{L}\geq 2)$, $\nu \ (\nu \simeq 0.01)$, and $\eta \ (\eta \simeq n/2)$ are the number of tree-based learners, mean depth of the tree-based learners, the shrinkage rate, and the sample fraction for each tree-based learner in training, respectively, which are given as hyper parameters. To update model $H$, we use the gradient boosting tree (GBT) method [Friedman, 2001].
$H$ is successively updated by the regression tree model [Breiman et al., 1984] $h_m$ using a greedy stagewise approach.\\
In lines $1$ to $3$ of Algorithm $1$, the model is initialized as $H_0(\bm{x}_i,z_i)$. Next, for each $m \ (m=1,2,\cdots, M)$, the pseudo-residual $z_m$ is calculated, as shown in line $6$ of Algorithm 1. Subsequently, in line $8$, the number of terminal nodes for $m$th tree-based learner $t_m$ is calculated as:
\begin{align*}
t_m = 2 + floor(u) \quad u \sim \exp(-u/(\bar{L}-2))/(\bar{L}-2),
\end{align*}
where $floor(\cdot)$ is the floor function and $\bar{L} \ (\bar{L}\geq 2)$ is the mean depth of the tree-based learners [Friedman and Popescu, 2008]. This random setting of the number of terminal nodes for each tree enables the production of trees of different sizes.
Then, a regression tree providing the disjoint terminal regions $R_{qm} \ (q=1, 2, \cdots, t_m; m=1, 2, \cdots, M)$ is fitted to the pseudo-residual $z_{im}$. 
In line $11$ of Algorithm 1, different optimal constants $\gamma_{qm}$ exist in each region. For calculations from line $9$ to $11$, we used the R package {\choosefont{pcr}rpart} [Therneau and Atkinson, 2022].
With these values, $H_m$ is updated, as shown in line $13$.
After generating $M$ regression trees, $K$ rule functions are constructed from them, as shown in line $16$.
Here, $K$ is the total number of rules generated from all trees, which can be calculated as:
\begin{align*}
K = \sum_{m=1}^M2(t_m-1)
\end{align*}
where $t_m$ denotes the number of terminal node in $m$th tree. 
In line $18$, the $K$ rules are combined.

{\footnotesize
\begin{spacing}{0.01}
\begin{figure}[H]
\label{alg1}
\begin{algorithm}[H]
    \caption{Rule generation}
    \algsetup{linenosize=\small}
\small
    \begin{algorithmic}[1]    
    \REQUIRE training set $\{y_i, (\bm{x}_i, z_i)\}_{i=1}^n$, number of tree-based learners $M$, mean depth of tree-based learners $\Bar{L}$, shrinkage rate $v$, and training sample fraction for each tree-based learner $\eta$ 
    \FOR{$i = 1$ to $n$}
    \STATE Set the initial model $H_0(\bm{x}_i, z_i) \leftarrow \bar{y}$
    \ENDFOR\\
    \FOR {$m=1$ to $M$}
    \FOR {$i=1$ to $n$}
    \STATE Compute the pseudo residual\\
    $\hspace{10pt} z_{im} \leftarrow \Tilde{y}_i - H_{m-1}(\bm{x}_i,z_i)$ 
    \ENDFOR\\
    \STATE Calculate the number of terminal nodes for the tree-based learner\\
    $\hspace{10pt}t_m = 2+floor(u) \hspace{1pt}$, where $u \sim \exp(-u/(\bar{L}-2))/(\bar{L}-2)$ \\
    \STATE Fit a regression tree to the pseudo residual $z_{im}$,  giving the 
    terminal regions $R_{qm} \quad (q=1,2, \cdots, t_m)$
    \FOR{$q=1,2, \cdots, t_m$}
    \STATE Estimate the value in region $R_{qm}$\\
    
    $\hspace{30pt} \hat{\gamma}_{qm} \leftarrow  \argmin_{\gamma} \sum_{\bm{x}_i \in R_{qm}}
    \sum_{i \in O} (z_{im}- (H_{m-1} + \gamma))^2$
    
     where $O \subset \{1, 2, \cdots, n \}$ and ($\left| O \right|$  = $\lfloor \eta \rfloor$ )\\
    \ENDFOR\\
    \STATE Update $H_m(\bm{x}_i,z_i) \leftarrow H_{m-1}(\bm{x}_i, z_i)+\nu \cdot \sum_{q=1}^{t_m}\hat{\gamma}_{t_m}I(\bm{x}_i\in R_{qm}) = H_{m-1}(\bm{x}_i, z_i)+\nu h_m(\bm{x}_i, z_i)$
    \ENDFOR\\
    \FOR{$m=1$ to $M$}
    \STATE Compose the rules $\{r_{k_m}(\bm{x}_i,z_i)\}_{{k_m}=1}^{K_m}$ from $h_m(\bm{x}_i,z_i)$
    \ENDFOR\\
    \STATE Collect the all rule set $\{r_{k_1}(\bm{x}_i,z_i)\}_{k_1=1}^{K_1}, \cdots,\{r_{k_M}(\bm{x}_i,z_i)\}_{k_M=1}^{K_M}$ as  $\{r_k(\bm{x}_i,z_i)\}_{k=1}^K \quad (K = \sum_{m=1}^MK_m)$
    \end{algorithmic}
\end{algorithm}
\end{figure}
\end{spacing}
}

\subsubsection{Rule ensemble and parameter estimation using group lasso}
3.2.2 is closely associated with the two advantages of the proposed method. First, the rule term function $r_k(\cdot)$ generated in $3.2.1$ is divided into rules related to treatment effects $r^*_{k^*}(\cdot)$ and others $r^\dagger_{k^\dagger}(\cdot)$, thereby indicating that the model in Eq. (5) contains the base functions relevant to the treatment effect and that of main effect, respectively. This enables the estimation of treatment effects for nonlinear relationships while considering the main effects. 
Second, to select the rules that contribute to the outcome, the proposed method uses group lasso [Yuan and Lin, 2006] to interpret the treatment effects based on the selected rules. The conventional Rulefit method uses lasso [Tibshirani, 1996] to prune the generated rules. 
In the case of lasso, if a rule is selected for only one of the two treatment groups, it does not specify whether that rule affects the outcomes. The necessity of this concept is referred to in Powers et al. [2018] as shared basis for both the target treatment group and control group.

Details of 3.2.2 are described in Algorithm 2. 
As mentioned in 2.2, a linear term is introduced in line $1$ and $6$ of Algorithm 2. In line $7$, the $r_k$ generated in $3.2.1$ is divided into rules related to the treatment effects $r^*_{k^*}$ and the others $r^\dagger_{k^\dagger}$, and the model in Eq. (5).
In line $7$, the $K$ rules are divided into $K^*$ rules for the treatment effects and $K^\dagger$ rules for the others. 
%
%
To estimate the parameters using group lasso, the group information of the rule terms is introduced as :
\begin{align}
\mathscr{C} = \{D_1,D_2,\cdots, D_{K^\dagger}, G_1, G_2, \cdots, G_{K^*}  \}
\label{gval}
\end{align}
where the singleton set of rules is related to the main effects $D_{k^\dagger}=\{k^\dagger\} (k^\dagger = 1, 2, \cdots, K^\dagger)$ and the set of two pairs $G_{k^*} = \{ (k^*, z=1), (k^*, z=0) \} \ (k^*=1,2,\cdots, K^*)$ includes the treatment effect.

The R package {\choosefont{pcr}grpreg} is used to estimate of the parameters and the hyper parameter $\lambda \ (\lambda>0)$ is selected by cross-validation using this package. The rule terms of the main effects and the linear term in Eq. (5) is the common term for both treatment groups, indicating that these parameters do not belong to group $h$; their regularization is treated as a traditional lasso.

Then, the regression parameters $\beta_0, \beta_{k^\dagger}, \alpha_j$, and $\bm{\beta}^{\ddagger}_{k^*}= ( \beta^{(a)}_{k^*}, \beta^{(c)}_{k^*})^T$ such that Eq. (9) is minimized. 
Here, $\hat{\beta_0}$ is the estimated intercept, $\hat{\beta}_{k^\dagger}$ is the estimated coefficients relevant to main effect, and $\hat{\alpha}_j$ is the estimated coefficients of linear term. Additionally, $\hat{\bm{\beta}}^{\ddagger}_{k^*}= ( \hat{\beta}^{(a)}_{k^*}, \hat{\beta}^{(c)}_{k^*})^T$ is the estimated coefficients relevant to HTE, where $\hat{\beta}^{(a)}_{k^*}$ and $\hat{\beta}^{(c)}_{k^*}$ are for target treatment group and control group, respectively.

{\footnotesize
\begin{spacing}{0.01}
\begin{figure}[H]
\begin{algorithm}[H]
    \caption{Rule ensemble and parameter estimation}
    \algsetup{linenosize=\small}
\small
    \label{alg2}
    \begin{algorithmic}[1]    
    \REQUIRE $\{y_i, (\bm{x}_i, z_i)\}_{i=1}^n$, estimated rules $\{r_k(\bm{x}_i, z_i)\}_{k=1}^K$, values of Winsorized margin $(\delta_j^-, \delta^+_j)\ (j=1,\cdots, p)$, $\lambda \ (\lambda >0)$
    \FOR {$i=1$ to $n$}
    \FOR {$j=1$ to $p$}
    \STATE $l_j(\bm{x}_i) = {\rm min}(\delta^+_j,{\rm max}(\delta_j^-,x_{ij}))$
    \STATE $l_j(\bm{x}_i) \leftarrow 0.4 \cdot l_j(\bm{x}_i)/std(l_j(\bm{x}_i))$
    \ENDFOR\\
    \ENDFOR\\
    \STATE Divide the rule terms $\{r_k(\bm{x}_i,z_i)\}_{k=1}^K$ into those related to treatment effects $\{r^*_{k^*}(\bm{x}_i)\cdot I(z_i \in V_{zk^*})\}_{{k^*}=1}^{K^*}$ and the other ones $\{r^\dagger_{k^\dagger}(\bm{x}_i)\}_{{k^\dagger}=1}^{K^\dagger}$\\
    \STATE Create group information to apply the group lasso\\
    $\hspace{10pt}\mathscr{C} = \{D_1,D_2,\cdots, D_{K^\dagger}, G_1, G_2, \cdots, G_{K^*}  \}$\\
    \STATE Estimate the coefficient vectors using group lasso\\
    \begin{align*}
    &(\hat{\beta}_0,\{\hat{\beta}_{k^\dagger} \}_{{k^\dagger}=1}^{K^\dagger},\{\hat{\alpha}_j\}_{j=1}^p, \{\hat{\beta}_{k^*}^\ddagger \}_{{k^*}=1}^{K^*})\\ 
    &= \underset{\bigl( \beta_0,\{\beta_{k^\dagger}\}_{{k^\dagger}=1}^{K^\dagger}, \{\alpha_j\}_{j=1}^p, \{\beta_{k^*}^\ddagger \}_{{k^*}=1}^{K^*}\bigr)} {\operatorname{argmin}}\frac{1}{2}\sum_{i=1}^N (y_i - F(\bm{x}_i, z_i))^2 
    + \lambda \left(\sum_{{k^\dagger}=1}^{K^\dagger} | \beta_{k^\dagger} | + \sum_{j=1}^p|\alpha_j |
    + \sqrt{2}\sum_{{k^*}=1}^{K^*} \| \beta_{k^*}^\ddagger \|^2_F \right)
    \end{align*}\\
    where $\beta_{k^*}^\ddagger = \bigl( \beta_{k^*}^{(a)}, \beta_{k^*}^{(c)}\bigr)$.
    \RETURN $\hat{\beta}_0,\{\hat{\beta}_{k^\dagger}\}_{{k^\dagger}=1}^{K^\dagger},\{\hat{\alpha}_j\}_{j=1}^p, \{\hat{\bm{\beta}}_{k^*}^\ddagger \}_{{k^*}=1}^{K^*}$
    \end{algorithmic}
\end{algorithm}
\end{figure}
\end{spacing}
}

\subsubsection{HTE calculation}
\label{step3}

In 3.2.2, we estimate each parameter of two treatment groups using common base function to calculate the HTE. 
From the model in Eq. (5), Eq. (2) in our proposed method for the target treatment group $z=1$ and the control group $z=0$ can be expressed as follows:
\begin{align}
\hat{\mu}_1(\bm{x}_i) &= \hat{\beta}_0 + \sum_{k^\dagger=1}^{K^\dagger} \hat{\beta}_{k^\dagger}r^\dagger_{k^\dagger}(\bm{x}_i) + \sum_{j=1}^p \hat{\alpha}_jl_j(\bm{x}_i)
+ \sum_{k^*=1}^{K^*}\hat{\beta}_{k^*}^{(a)}r_{k^*}^*(\bm{x}_i) \quad {\rm and}
\label{temodel1} \\
\hat{\mu}_0(\bm{x}_i) &= \hat{\beta}_0 + \sum_{k^\dagger=1}^{K^\dagger} \hat{\beta}_{k^\dagger}r^\dagger_{k^\dagger}(\bm{x}_i) + \sum_{j=1}^p \hat{\alpha}_jl_j(\bm{x}_i)
+  \sum_{k^*=1}^{K^*}\hat{\beta}_{k^*}^{(c)}r_{k^*}^*(\bm{x}_i).
\label{temodel2}
\end{align}
Regardless of the treatmet, the main effects and linear terms of both treatment groups in Eq. (11) and Eq. (12) are the same.  
From Eq. (11) and Eq. (12), the HTE in Eq. (1) is calculated as follows:  
\begin{align}
\hat{\tau}(\bm{x}_i) =& \left( \hat{\beta}_0 + \sum_{{k^\dagger}=1}^{K^\dagger} \hat{\beta}_{k^\dagger}r_{k^\dagger}(\bm{x}_i) + \sum_{j=1}^p \hat{\alpha}_jl_j(\bm{x}_i)
+ \sum_{k^*=1}^{K^*}\hat{\beta}_{k^*}^{(a)}r_{k^*}^*(\bm{x}_i)  \right) \nonumber\\
&- \left( \hat{\beta}_0 + \sum_{{k^\dagger}=1}^{K^\dagger} \hat{\beta}_{k^\dagger}r_{k^\dagger}(\bm{x}_i) + \sum_{j=1}^p \hat{\alpha}_jl_j(\bm{x}_i)
+ \sum_{{k^*}=1}^{K^*}\hat{\beta}_{k^*}^{(c)}r_{k^*}^*(\bm{x}_i) \right) \nonumber\\
=& \sum_{{k^*}=1}^{K^*}\hat{\beta}_{k^*}^{(a)}r_{k^*}^*(\bm{x}_i)
-  \sum_{{k^*}=1}^{K^*}\hat{\beta}_{k^*}^{(c)}r_{k^*}^*(\bm{x}_i)\nonumber\\
=& \sum_{{k^*}=1}^{K^*} \bigl(\hat{\beta}_{k^*}^{(a)}-\hat{\beta}_{k^*}^{(c)}\bigr)r_{k^*}^*(\bm{x}_i).
\label{hte0}
\end{align}
This indicates that the HTE can be calculated using Eq. (1) with terms for each treatment arm. Therefore, the HTE of the proposed method can be estimated using the difference in the predicted values of each treatment arm, considering the main effects of the estimation.

\section{Numerical simulation}
\label{simu}

Numerical simulations are conducted to evaluate the performance of the proposed method.
We expected the results of the proposed method to be equivalent to the results of the compared methods.
In this section, we explain the simulation design and present the results.

\subsection{Simulation design}

First, we generated the covariate matrix $\bm{X} = (x_{ij})\:(i=1, 2, \cdots, n; j=1, 2, \cdots, p)$. $x_{ij}$ was randomly distributed from $N(0,1)$, where $N$ is a normal distribution.
Our setting is a two-armed randomized controlled trial; therefore, we set the treatment group variable as $z_i =  \{ 1, 0 \}$ as the treatment arm, where $z_i=1$ and $z_i=0$ signify the target treatment group and the control group, respectively. The treatment group indicator $z_i$ was generated based on Bernoulli distribution $z_i \sim B(0.5)$.

Using $\bm{x}_i$ and $z_i$, the outcome variable was randomly generated as:
\begin{align*}
y_i = \psi(\bm{x}_i) + \left(z_i - \frac{1}{2} \right)\tau(\bm{x}_i)+\epsilon_i
\end{align*}
where $\psi: \mathbb{R}^p \mapsto \mathbb{R}$ is the true effect related to the outcome of the covariates $\bm{x}_i$ and $\tau(\bm{x}_i)$ is that of the HTE. The error distribution $\epsilon_i$ follows the normal distribution $N(0,0.25)$.
We generated the training and the test data using the same settings and sample sizes.
To compare the performance of the proposed method and the compared methods, the simulation was conducted with various factors. 
The total pattern of the simulation was $2$ (Factor 1) $\times$  $3$ (Factor 2) $\times$ $4$ (Factor 3) $\times$ $4$ (Factor 4) = $96$.

We present the factors of the simulation settings below.

\noindent
{\bf Factor 1: Sample Size}\\
The sample size $n$ is $600$ and $1000$ to examine the influence of the number $n$.\\

\noindent
{\bf Factor 2: Covariate Variables}\\
The number of variable $p$ is set $100$, $200$ and $400$ to examine the influence of the number of $p$. \\

\noindent
{\bf Factor 3: Patterns of $\psi(\bm{x})$}\\
$\psi(\bm{x}_i)$ is a function that expresses the main treatment effects. We set four different settings, as listed in the second column of Table 1. Scenario $1$ to $4$ and $5$ to $8$ assume linear functions. Scenario $9$ to $12$ were generated from the nonlinear function by the indicator function, while Scenario $13$ to $16$ were generated by $\sin$ function and exponential function.\\

\noindent
{\bf Factor 4: Patterns of $\tau(\bm{x})$}\\
$\tau(\bm{x}_i)$ generates data relevant to the HTE. We set other $4$ different settings shown in the third column from the left of Table 1. Scenario $1,5,9$ and $13$ are the combinations of linear and quadratic functions. Scenario $2, 6, 10$, and $14$ are indicator functions that assume the quantitative data. Scenario $3, 7, 11$, and $15$ are based on $\sin$ function and exponential function. Scenario $4, 8, 12$, and $16$ assume no treatment effects.

To evaluate the performance accuracy, we used three different evaluation indices. The first is the mean squared error (MSE) calculated as  
\begin{align*}
MSE = \frac{1}{n}\sum_{i=1}^n\bigl(\tau^*(\bm{x}_i) - \hat{\tau}(\bm{x}_i)\bigr)^2 
\end{align*}
where $\tau^*(\bm{x}_i)$ is the true HTE value and $\hat{\tau}(\bm{x}_i)$ is the estimated HTE value. The second evaluation index is the relative bias (RBias) against the true HTE, given as 
\begin{align*}
RBias = \frac{1}{n}\sum_{i=1}^n
\frac{\bigl(\tau^*(\bm{x}_i) - \hat{\tau}(\bm{x}_i)\bigr)}{\tau^*(\bm{x}_i)}.
\end{align*}
The third evaluation index is Spearman's correlation coefficient between the true HTE and the estimated HTE. In practical situations, subgroups are detected by subjects ordered based on estimated HTE [Tian et al., 2014], therefore, we add this evaluation.

We compared the proposed method to $4$ different tree-based approach methods: Causal Forest [Wager and Athey, 2018], Bayesian additive regression trees (BART) [Hill, 2011], Causal MARS (Multivariate Adaptive Regression Spline) [Powers et al., 2018], and Pollinated Transformed Outcome (PTO) forest [Powers et al., 2018]. In the simulation, we used {\choosefont{pcr}Rstudio version}. 
We used the R packages {\choosefont{pcr}grf} [Tibshirani et al., 2022] for the Causal Forest, {\choosefont{pcr}bartCause} [Hill, 2011] for BART, and {\choosefont{pcr}causalLearning} [Powers et al., 2022] for the Causal MARS and PTO forest.

\begin{table}
\begin{center}
\caption{Scenarios of the simulation }\label{simpat}
\begin{tabular}{r|l|l} 
\hline
$\rm{Scenario}$ & $\psi(\bm{x})$ & $\tau(\bm{x})$ \\\hline
$1$ & $x_1x_2$ & $2(x_2)+x_3^2+x_5x_6+x_8^2$\\
$2$ & $x_1x_2$ & 
$2 + 0.3I(x_4>-3) - 4I(x_5>0) + 0.7I(x_7<1)$\\
$3$ & $x_1x_2$ & $3\sin(x_1x_5)^2 + 5\exp(x_8+x_3)$\\
$4$ & $x_1x_2$ & $0$\\\hline
$5$ & $x_1+x_3-x_5$ & $2(x_2)+x_3^2+x_5x_6+x_8^2$\\
$6$ & $x_1+x_3-x_5$ & 
$2 + 0.3I(x_4>-3) - 4I(x_5>0) + 0.7I(x_7<1)$\\
$7$ & $x_1+x_3-x_5$ & $3\sin(x_1x_5)^2 + 5\exp(x_8+x_3)$\\
$8$ & $x_1+x_3-x_5$ & $0$\\\hline
$9$ & $0.5I(x_1>-1) -1.4I(x_3>0)$ & $2(x_2)+x_3^2+x_5x_6+x_8^2$\\
$10$ & $0.5I(x_1>-1) -1.4I(x_3>0)$  & 
 $2 + 0.3I(x_4>-3) - 4I(x_5>0) + 0.7I(x_7<1)$\\
$11$ & $0.5I(x_1>-1) -1.4I(x_3>0)$  & 
$3\sin(x_1x_5)^2 + 5\exp(x_8+x_3)$\\
$12$ & $0.5I(x_1>-1) -1.4I(x_3>0)$  &  $0$\\\hline
$13$ & $3\sin(x_4 + x_5)^2 -0.2\exp(x_7)$ & 
$2(x_2)+x_3^2+x_5x_6+x_8^2$\\
$14$ &  $3\sin(x_4 + x_5)^2 -0.2\exp(x_7)$ & 
 $2 + 0.3I(x_4>-3) - 4I(x_5>0) + 0.7I(x_7<1)$\\
$15$ & $3\sin(x_4 + x_5)^2 -0.2\exp(x_7)$ & 
$3\sin(x_1x_5)^2 + 5\exp(x_8+x_3)$\\
$16$ & $3\sin(x_4 + x_5)^2 -0.2\exp(x_7)$ & $0$\\\hline
\end{tabular}
\end{center}
\end{table}

\subsection{Simulation results}
The results are presented in Figure 1 to Figure 6. 
First, we explain the results of MSE. 
Figure 1 is the MSE of $n=600$ and Figure 2 is that of $n=1000$. The horizontal axis represents the number of variable $p$ and the vertical axis representes the MSE value. Both figures are drawn by method, and each figure is plotted by the Scenario in Table 1. 
Overall, the results of the proposed method were better, particularly in cases where the settings of the true HTE $\tau$ were more complicated nonlinear functions.
Additionally, the proposed method was stable regardless of the value of $n$ and $p$, whereas the MSE of Causal Forest and BART increased depending on the value of $p$. 
Now, we observe the results for the scenarios.
In Scenario $1, 4, 9,$ and $13$ of $n=600$ (the leftmost column of Figure 1), whose setting of the true HTE is a combination of linear and quadratic functions, the MSE values of Causal MARS were found to be the least. The proposed method was superior to the other compared methods, except for Causal MARS in $p=200$ and $400$ in Scenario $13$. 
For Scenario $2, 6, 10,$ and $14$ (the second column from the left of Figure 1), whose setting of the HTE is piecewise constant, the proposed method and PTO forest were superior to the other methods in Scenario $6$ and $10$. For $p=100$ and $200$ in Scenario $2$, the MSE values of the proposed method were smaller than those of PTO forest, although it was slightly increased in $p=400$. The MSE of PTO forest was smaller in Scenario $14$. 
In Scenario $3, 7, 11$ and $15$ of $n=600$ (the second column from the right of Figure 1), whose setting of HTE is a combination of $\sin$ function and exponential function, the proposed method was not smaller than Causal MARS and PTO forest; however, the difference among them was rather slight compared to that of the Causal Forest and BART. Regarding the trend of the MSE values, the proposed method, Causal MARS, and PTO forest  remained as the value of $p$ increased. The MSE values of the Causal Forest and BART increased significantly as the number of variable increased.
In Scenario $4, 8, 12,$ and $16$ (the rightmost column of Figure 1), MSE of the proposed method and BART were estimated as $0$, and Causal Forest also estimated nearly to the true $\tau$ value. On the other hand, Causal MARS and PTO forest in Scenario $4$ and $8$ estimated the presence of the treatment effect, and the trend depending on $p$ was unstable. 
In the case of $n=1000$, Scenario $1, 4, 9,$ and $13$ (the leftmost column of Figure 2), Causal MARS was also better than the other methods, however, the proposed method was superior to the PTO forest in all Scenarios. 
In Scenario $2, 6, 10,$ and $14$ of $n=1000$  (the second column from the left of Figure 2), the proposed method was superior to the other methods in Scenario $2$, and was almost the same as PTO forest in Scenario $6$ and Scenario $10$. In Scenario $14$, PTO forest was better than the proposed method; however, the difference between these two methods was closer than that in $n=600$ of the same scenario.
In all scenarios, the trend of the MSE with respect to the number of $p$ showed a tendency similar to that in the case of $n=600$. The MSE values of the proposed method, Causal MARS and PTO forest were not influenced by the number of $p$, while those of Causal Forest and BART increased. 

Next, we compare the results for the relative bias between $\tau^*$ and $\hat{\tau}$ values displayed in Figure 3 and Figure 4.
Figure 3 shows the bias of $n=600$.
The bias in Scenario $4, 8, 12$ and $16$ does not create a plot because $\tau^*$ is set to $0$.
Figure 3 shows the bias of $n=600$. Each Scenario had plots of $p=100, 200$ and $400$.
Almost all results of the median value were positive for all methods.
The proposed method, Causal MARS, and PTO forest were almost stable even when $p$ increased. However, BART and Causal Forest showed a larger bias as $p$ increased. These results showed that the proposed method, Causal MARS and PTO forest were in estimation regardless of the number of $p$, whereas BART and Causal Forest increased the bias by the influence of the number of $p$. 
We examined the results of each scenario. 
In Scenario $1, 5, 9$, and $13$, the setting of the true HTE is a combination of linear and quadratic functions, the median values of Causal MARS were close to the true $\tau$ value, and PTO forest includes bias $0$. 
In Scenario $2, 6, 10$ and$14$, where setting of the HTE is set as piecewise constant, the median of PTO forest were nearly equal to $0$. The proposed method performed better next to PTO forest. However, Causal Forest and BART were more than $0.5$ at $p=400$.
In Scenario $3, 7, 11$, and $15$, the setting of the HTE is a combination of $\sin$ function and exponential function, the median of almost all methods was close to $0$. The results of the proposed method were slightly higher than $0$, whereas the range of the results was narrower than those of Causal Forest and BART. 
Figure 4 shows the results for $n=1000$. 
The tendency for $n=1000$ was similar to that for $n=600$ in all the scenarios.

Finally, we show the results of the correlation between the estimated treatment effect $\hat{\tau}$ and the true treatment effect $\tau^*$. 
The results were drawn in Figure 5 and Figure 6.
Figure 5 presents the results for $n=600$. 
The proposed method and the other compared methods were stable in almost all scenarios, regardless of $p$ except for BART. The correlation of BART was affected by the value of $p$.
Causal MARS had the highest correlation among the methods in Scenario $1, 5, 9$, and $13$, and the correlation of the proposed method was higher than that of PTO forest. 
In Scenario $2, 6$ and $14$, the proposed method had the higher correlation than the compared methods except $p=100$ in Scenario $6$. 
In Scenario $3, 7, 11$ and $15$, the correlation of all methods were close to $0$, except for BART at $p=400$. 
The results of $n=1000$ are presented in Figure 6. 
The overall correlation was slightly higher than that at $n=600$; however, the trends of the results for $n=1000$ were similar to those for $n=600$. 

From these results, in terms of MSE, the proposed method was found to be stable as the number of variable increased, while Causal Forest and BART affected the increase in the number of variable. These trends were confirmed by the results of RBias and correlation coefficients. Moreover, in most scenarios of the true treatment effects that comprise threshold functions, MSE of the proposed method showed better performance than the other compared methods and was better or almost close to MSE of Causal Mars and PTO forest for scenarios where the true treatment follows a combination of $\sin$ and $\exp$ function. The tendencies of the results for Rbias and correlation coefficients were similar to those of MSE. These results confirmed the estimation performance of the proposed method in non-linear structure. However, in the scenarios of a combination of linear and quadratic function in true treatment, Causal MARS was better than the proposed method, while the proposed method yielded better results than the other four compared methods. Causal MARS captured the structure of the linear tendency; however, the estimation performance was inferior for nonlinear structure. We confirmed that the proposed method performed well when the true treatment followed nonlinear structure and maintained the estimation accuracy as the near-linear structure.

\begin{figure}[htbp]
\begin{center}
\includegraphics[scale=0.41]{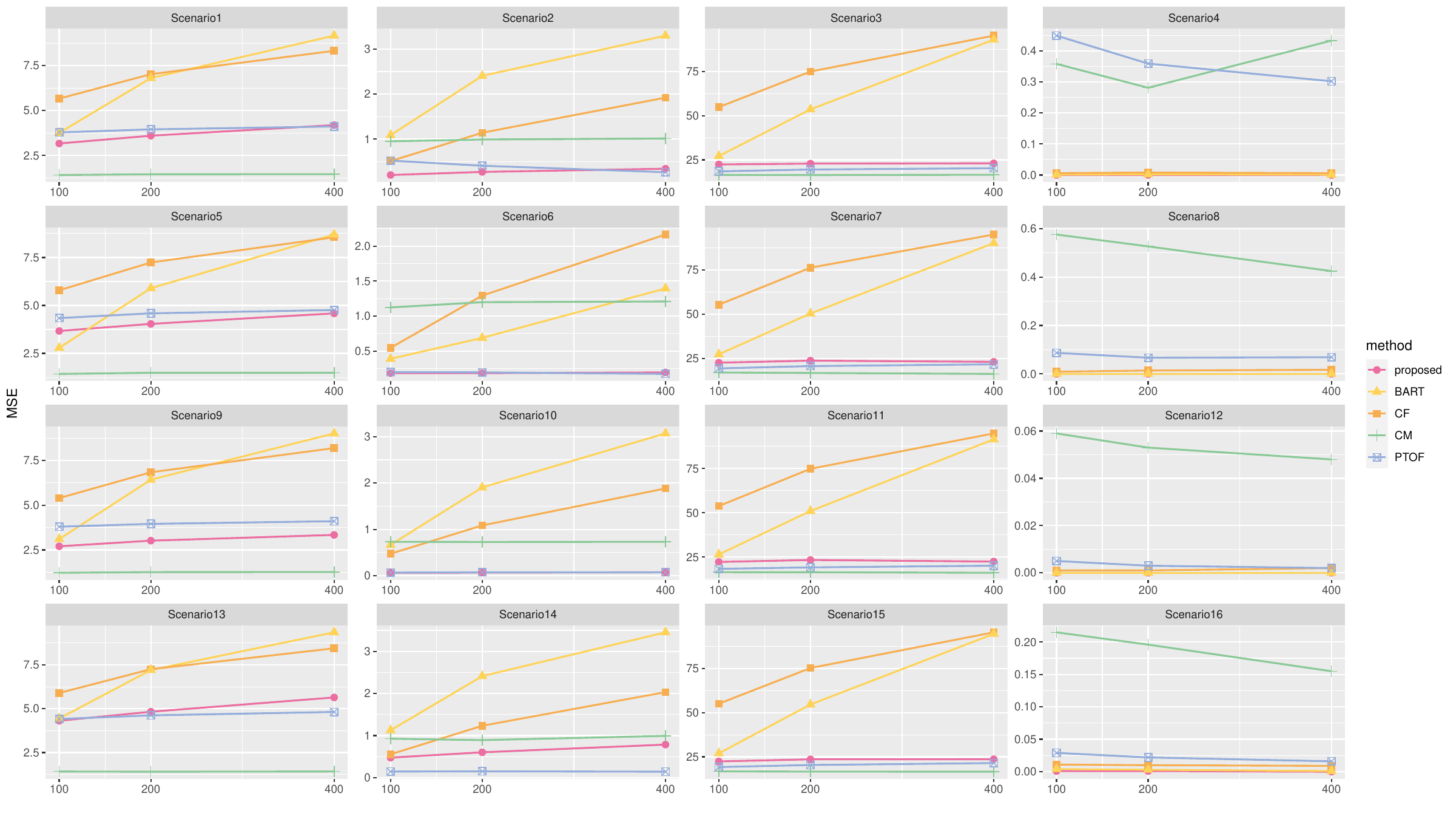}
\caption{Plots of MSE in $n=600$. The horizontal axis is the number of covariate variable and the vertical axis is MSE.}
\label{mse600}
\end{center}
\end{figure}

\begin{figure}[htbp]
\begin{center}
\includegraphics[scale=0.41]{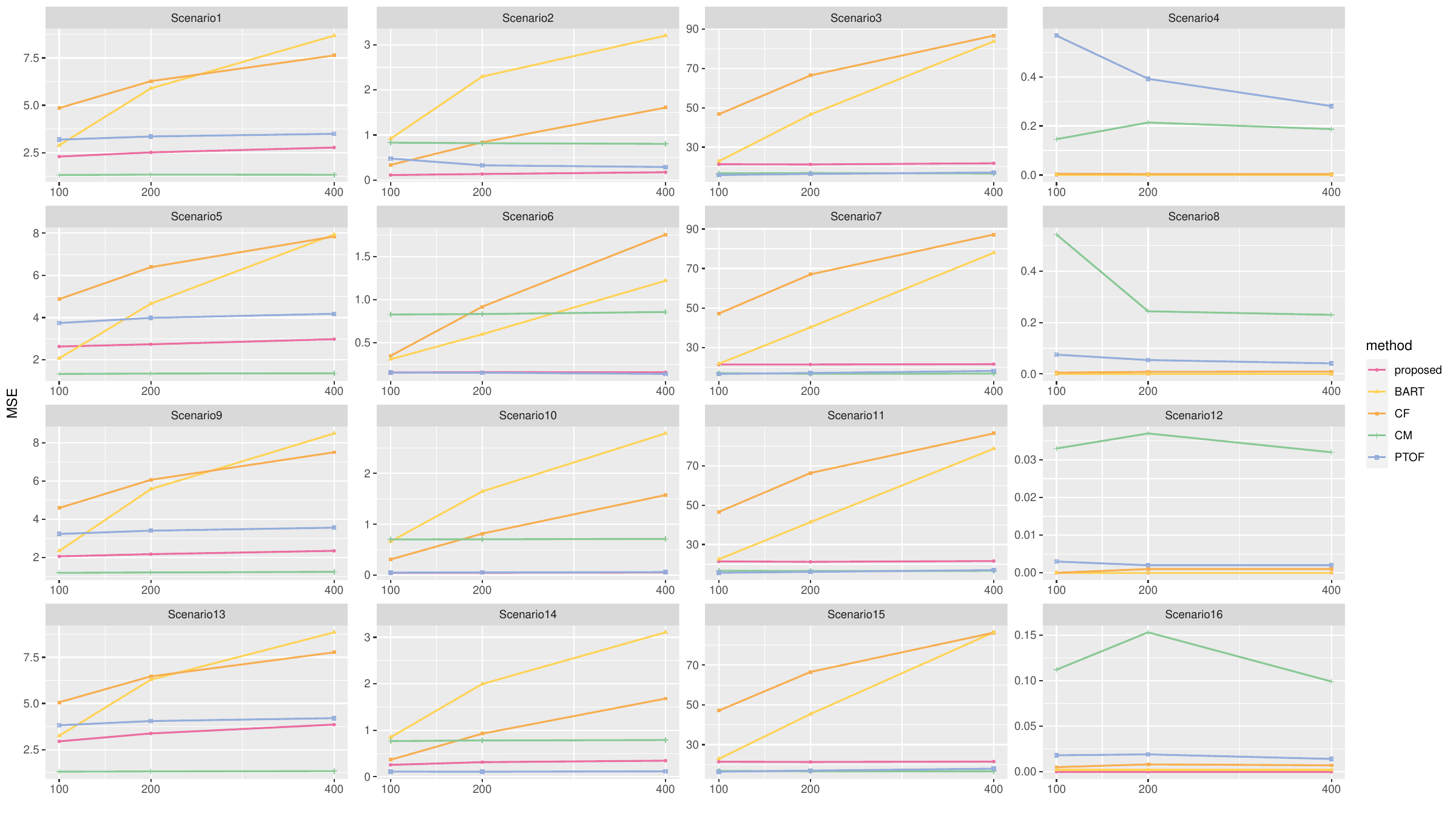}
\caption{Plots of MSE in $n=1000$. The horizontal axis is the number of covariate variable and the vertical axis is MSE.}
\label{mse1000}
\end{center}
\end{figure}

\begin{figure}[htbp]
\begin{center}
\includegraphics[scale=0.41]{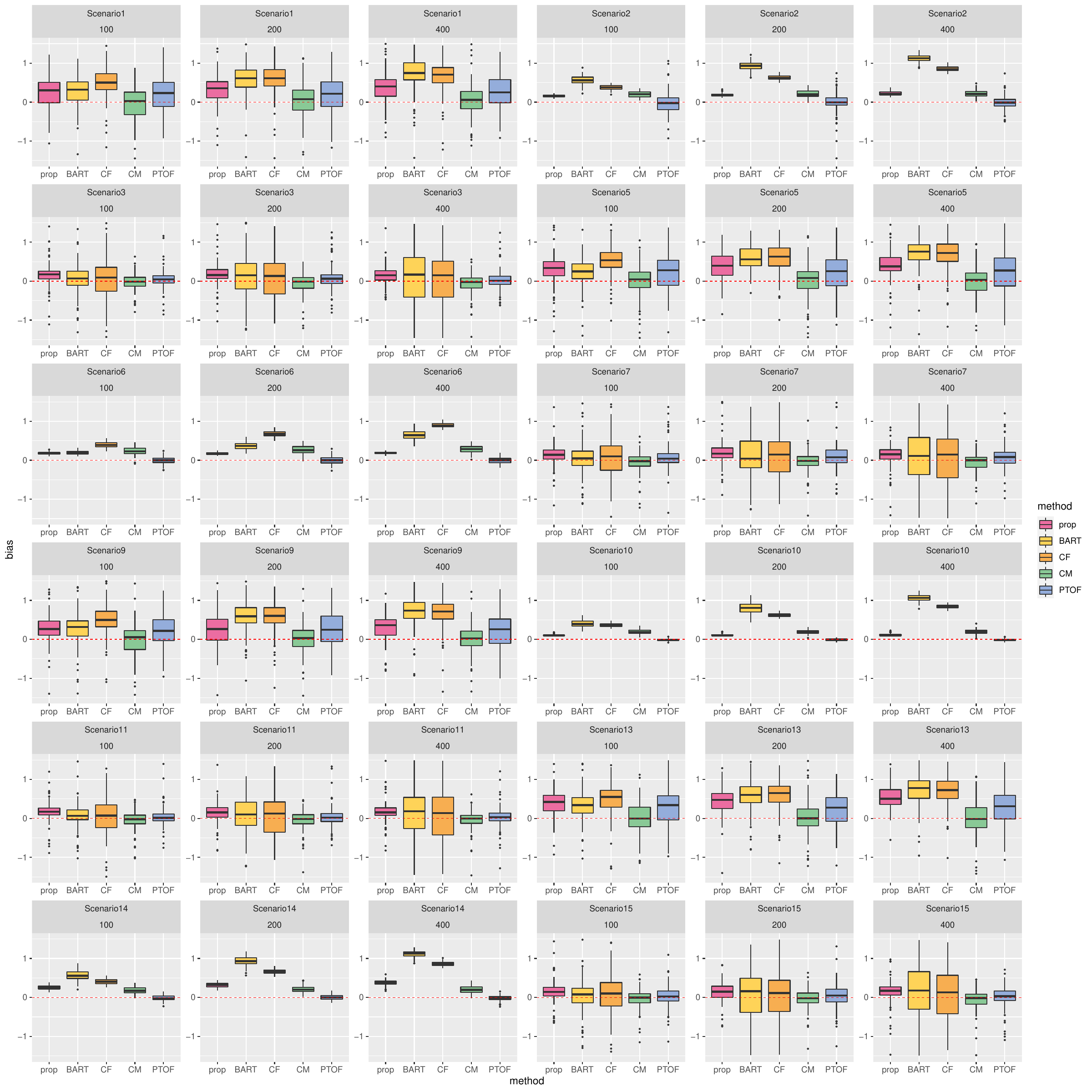}
\caption{Plots of bias in $n=600$. Each Scenario is depicted by $p =100, 200$ and $400$. The horizontal axis is method and the vertical axis is bias. Scenario $4, 8, 12,$ and $16$, are excluded due to the bias cannot be calculated by $\tau(x)=0$.}
\label{biasn600}
\end{center}
\end{figure}

\begin{figure}[htbp]
\begin{center}
\includegraphics[scale=0.41]{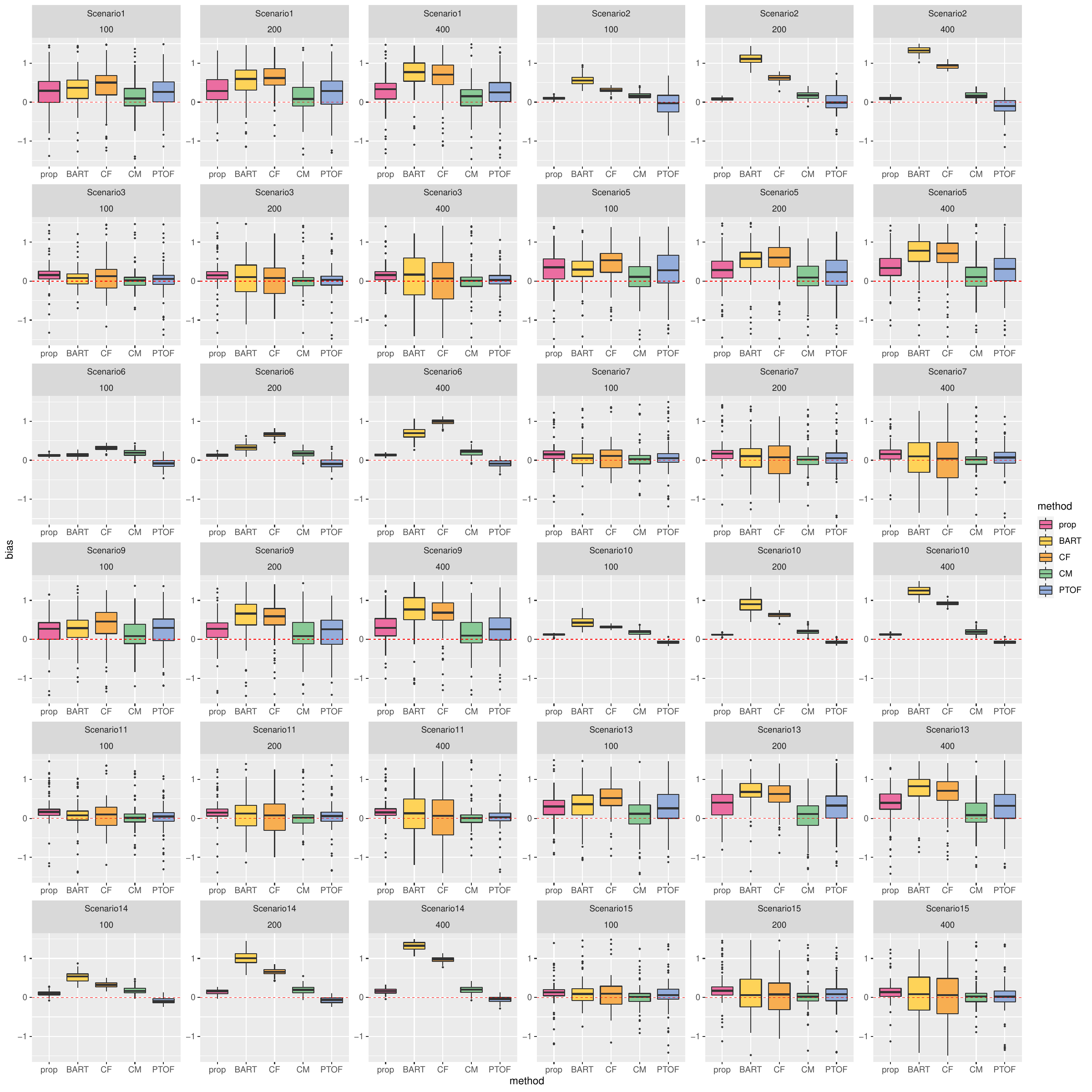}
\caption{Plots of bias in $n=1000$. Each Scenario is depicted by $p = 100, 200$ and $400$. The horizontal axis is method and the vertical axis is bias. Scenario $4, 8, 12,$ and $16$, are excluded due to the bias cannot be calculated by $\tau(x)=0$.}
\label{biasn1000}
\end{center}
\end{figure}

\begin{figure}[htbp]
\begin{center}
\includegraphics[scale=0.41]{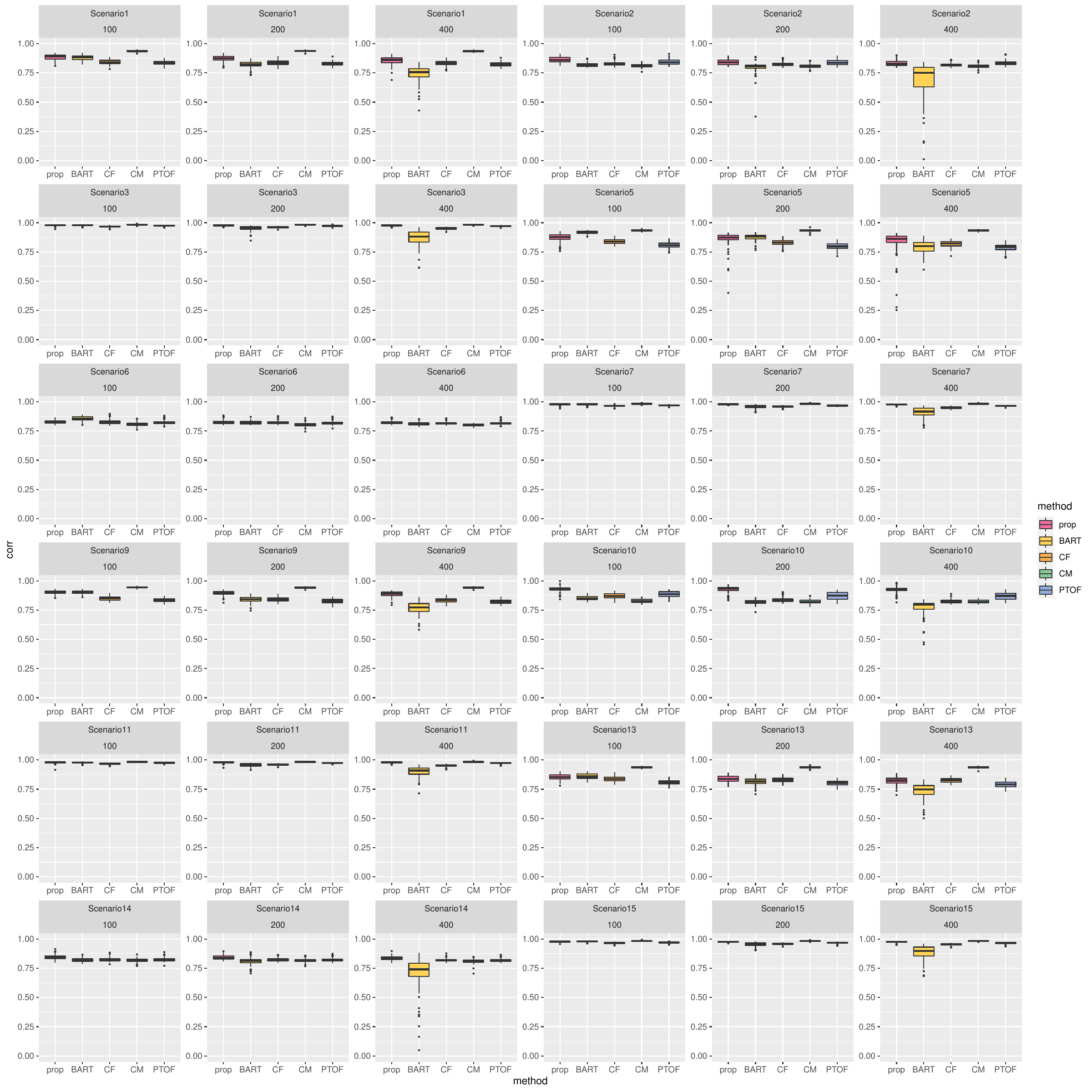}
\caption{Plots of correlation in $n=600$. Each Scenario is depicted by $ p =100, 200$ and $400$. The horizontal axis is method and the vertical axis is correlation. Scenario $4, 8, 12,$ and $16$, are excluded due to the bias cannot be calculated by $\tau(x)=0$.}
\label{corn600}
\end{center}
\end{figure}

\begin{figure}[htbp]
\begin{center}
\includegraphics[scale=0.41]{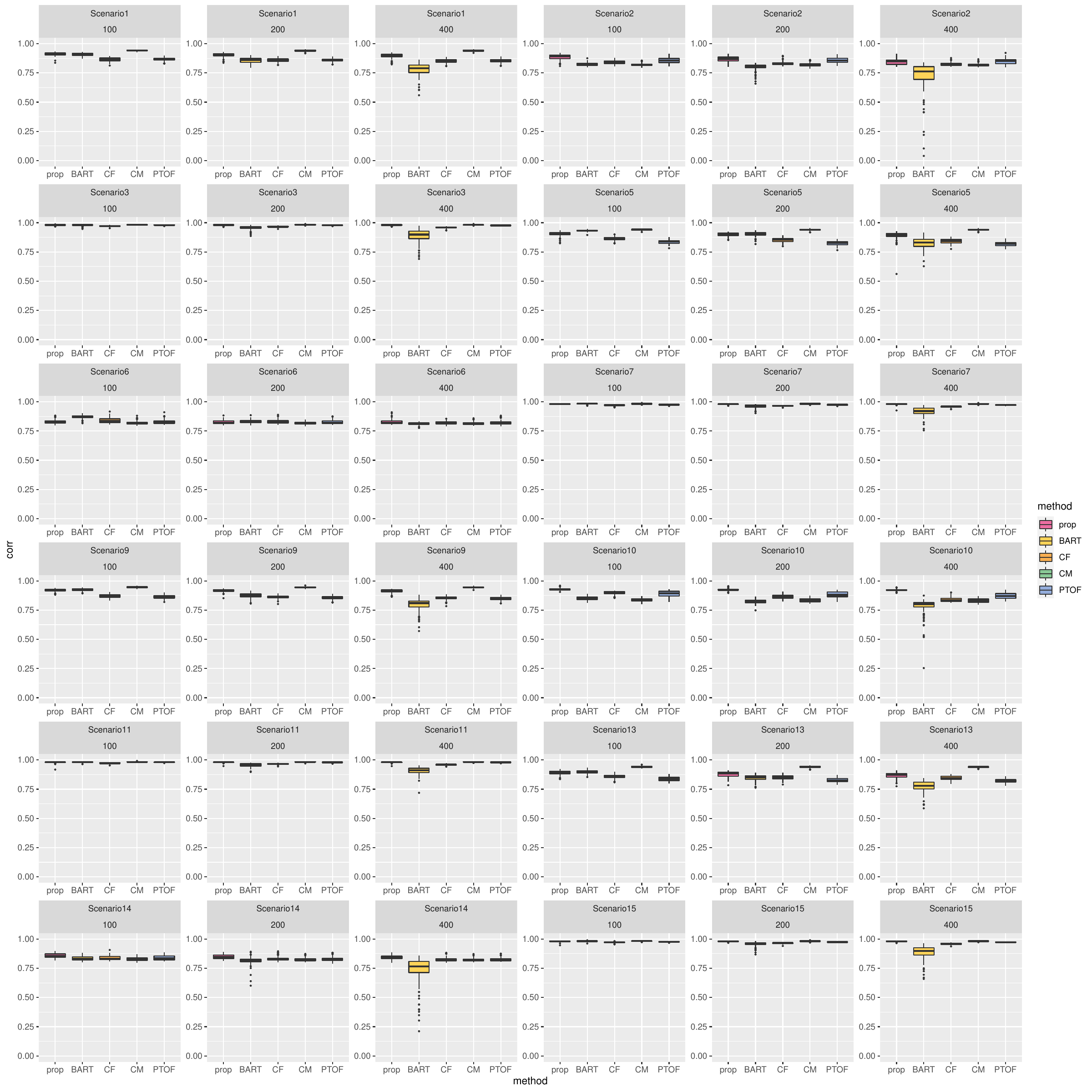}
\caption{Plots of correlation in $n=600$. Each Scenario is depicted by $p = 100, 200$ and $400$. The horizontal axis is method and the vertical axis is correlation. Scenario $4, 8, 12,$ and $16$, are excluded due to the bias cannot be calculated by $\tau(x)=0$.}
\label{corn1000}
\end{center}
\end{figure}

\newpage

\section{Real data application}
\label{appl}
In this section, we demonstrate the usefulness of the proposed method by applying it to actual clinical study data named AIDS Clinical Trials Group Study 175 (ACTG 175) [Hammer et al., 1996] from the package {\choosefont{pcr}speff2trial} [Juraska et al., 2022] in the {\choosefont{pcr}R software}. 
In this double-blind randomized study, $2139$ subjects infected with human immunodeficiency virus type 1 (HIV-1) at $200$ to $500$ per mL CD4 cell counts  were randomly assigned to one of the $4$ arms: zidovudine with didanosine, zidovudine with zalcitabine, zidovudine only, and didanosine only.
We selected $522$ subjects in the zidovudine plus zalcitabine group as the target treatment group and $532$ subjects in the zidovudine only group as the control group. The outcome was defined as the difference in CD4 cell counts at 20 $\pm$ $5$ weeks from their baseline. The $15$ covariates are selected in Table 2. 
For the hyper parameters of the proposed method, 
the number of trees was set to $M=400$, the shrinkage rate to $\eta = 0.25$, and the mean depth of each tree-based function to $\bar{L} = 2$. 

\begin{table}[H]
\begin{center}
\caption{Selected covariates at baseline in real data application.}\label{cov15}
\scalebox{0.8}{
\begin{tabular}{ll} 
\hline
variable name & description  \\\hline
age & age in years \\
wtkg & weight in kg \\ 
karnof & Karnofsky score, a scale of $0-100$ \\
preanti & the number of days of antiretroviral therapy previously received \\
cd40 & CD4  cell count at baseline, cells/mm$^3$ \\
cd80 & CD8  cell count at baseline, cells/mm$^3$ \\
hemo & hemophilia, $0=$ no, $1=$ yes\\
homo & homosexal activity, $0=$ no, $1=$ yes\\
drugs & history of intravenous drug use,  $0=$ no, $1=$ yes\\
oprior & non-zidovudine antiretroviral therapy prior to intiation of study treatment, $0=$ no, $1=$ yes\\
zprior & zidovudine use prior to treatment initiation, $0=$ no, $1=$yes \\
race & $0=$ white, $1=$ non-white\\
gender & $0=$ female, $1=$ male\\
str2 &  antiretroviral history, $0=$ naive, $1=$ exprerienced\\
symptom & symptom indicator, $0=$ asymptomatic, $1=$ symptomatic\\\hline
\end{tabular}
}
\end{center}
$\quad \quad \quad \quad \quad \quad \quad$ 
\end{table}

Based on the application results, we demonstrated the estimation results, and the obtained rules of the proposed method.

For the estimation results, we ordered the estimated HTE in an ascending sequence and divided them into three equal portions: low, middle and high.
If the HTE is properly estimated, the mean of the outcome will be low, middle, and high in the low group, middle group and high group, respectively.
The procedure used to divide the groups is described below.
First, the sample ID was arranged in ascending order of the estimated HTE. The ordered sample ID was then divided into three groups: low for $351$ subjects, middle for $351$ subjects, and large for $352$ subjects. 
The mean and standard error of the outcome for each treatment group were then calculated for each ordered group.
If HTE is properly estimated, the low group will expect that the difference in the mean of the outcome between the two treatment groups to be smaller than that in the middle and high groups. Conversely, the difference in the mean of the outcome in the high group will be expected to be larger than that in the other two groups. 
Next, we show the results of the three groups ordered by treatment arm in Figure 7. 
The two bars on the left are the results for the small group, the middle bars are those for the middle group, and the right bars are the results for the high group. The green and pink bars represent the target treatment group and the control group, respectively. The differences between the treatment arms mostly increased in the high group. 
Therefore, we confirmed that the estimated results of the proposed method exhibited a trend.

Additionally, we calculated the rule importance and its support to observe the subgroups of the data [Friedman and Popescu, 2008]. The advantage of the RuleFit method is its rule-based interpretability, and the conventional RuleFit method evaluates the importance of the rule and linear terms to the coefficient values. 
We focused on the rule importance [Friedman and Popescu, 2008] of rule terms related to HTE. 
The $k^*$th importance $Q_{k^*}$ for rules of the proposed method can be calculated as follow
\begin{align*}
Q_{k^*} = \bigl( |\hat{\beta}^{(a)}_{k^*}-\hat{\beta}^{(c)}_{k^*}| \cdot \sqrt{o_{k^*}(1-o_{k^*})} \bigr)
\end{align*}
where $o_{k^*}$ is the support of the rule importance for $r^*_{k^*}$.
Support indicates the percentage of subjects who meet the base function. The support in $k$th rule can be computed as 
\begin{align*}
o_{k^*}=\frac{1}{n}\sum_{i=1}^n r^*_{k^*}(\bm{x}_i).
\end{align*}

\begin{figure}[htbp]
\begin{center}
\includegraphics[scale=0.63]{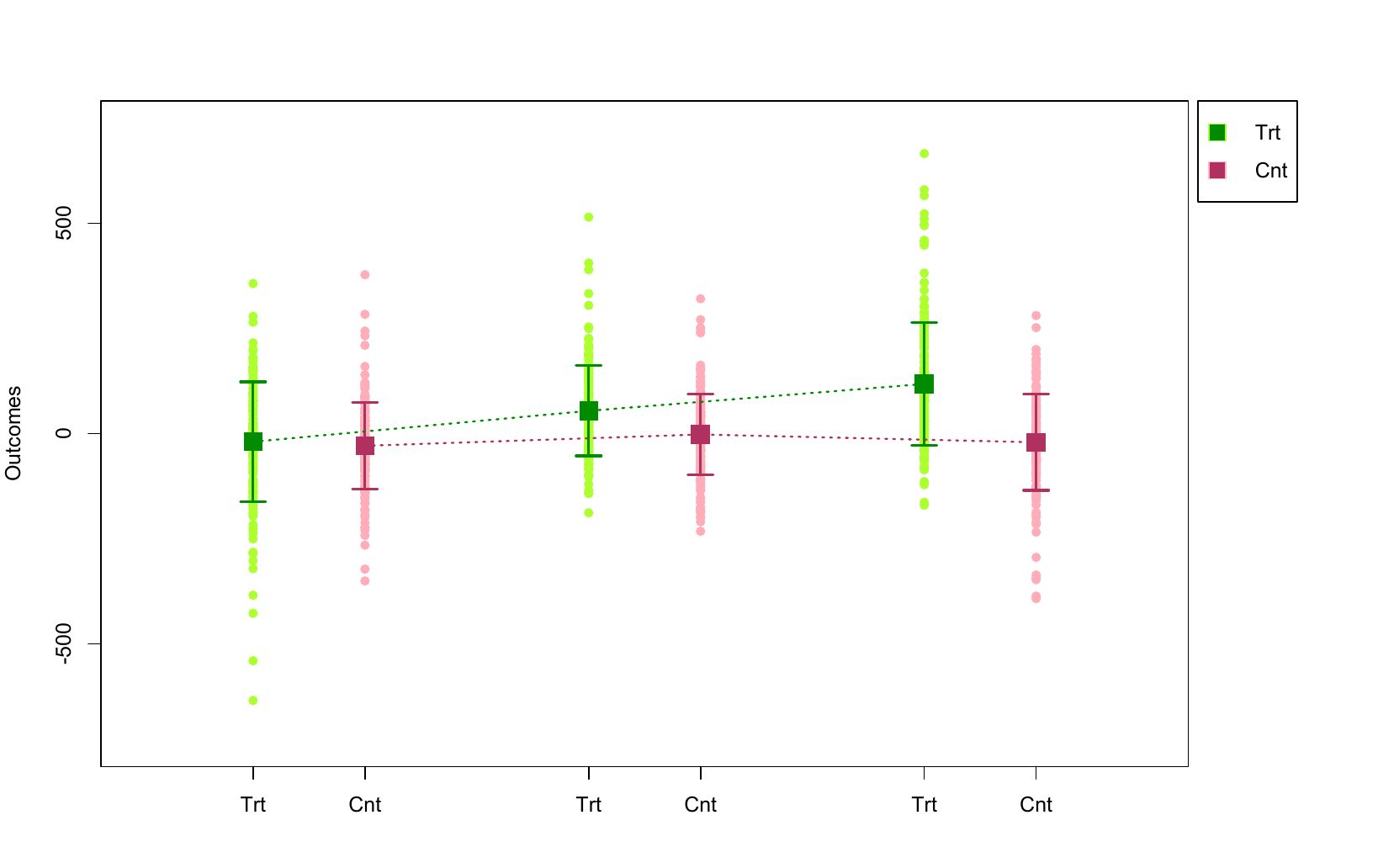}
\caption{Plots of the distribution of the outcome, its mean and its standard error by each arm in three group: low, middle, and high. The vertical axis is the outcome and the horizontal axis depicts each treatment. The green bar and plot is treatment group, and the pink bar and plot is control group. The left green and pink bars are the low group, whose estimated HTE were smaller in order. The middle bars and plots were the middle group, and the right green and pink bars and plots are the high group. }
\label{resmeth}
\end{center}
\end{figure}

In this application, $25$ rules were chosen for estimating HTE. 
Figure 8 shows the rule importance on the left and its support on the right.
On the left side of Figure 8, rules with high rule-importance values are indicated by pink bars.
The pink bar on the right side of Figure 8 indicates support values more than $0.7$. 


Furthermore, the proposed method can represent the characteristics of the subgroups relevant to the treatment effect as rules. Furthermore, we depicted a distribution of estimated HTE of $25$ rules and its support values in Figure 9. The vertical axis represents the estimated HTE for each rule and the horizontal axis represents support value corresponding to each rule. From this plot, we can observe the overall trend in the results. The HTE values of each rule were relatively located between $0$ and $15$, and their support values ranged widely.  In particular, rule \#$1$, for example, was high HTE value, whereas the support value was small, thereby indicating that the subgroup that fits rule \#$1$ did not meet less than $10$ \% of this data, although the HTE for this rule was high.
In contrast, the HTE of rule \#$8$ was approximately $24$, and its support value was approximately  $0.85$. 
Although the HTE was not high, $85$\% of the subjects belonged to this subgroup, thereby indicating that this subgroup was suitable as the subjects in this study.
To select the subgroups, we chose top $8$ rules in rule importance were selected.
The rules depicted by red point in Figure 9 are listed in Table 3. 
The HTE of $8$ rules was positive; therefore, the subgroups that met these rules were more effective in zidovudine and didanosine combination therapy than in zidovudine-only therapy. This result was consistent with ACTG 175 study results [Hammer et al., 1996, Saravolatz et al., 1996].  
When seeing the context, the rule ''wtkg$<102.4$ \& cd40 $<147$ ''
indicates that less than $102.4$ kg and less than $147$ cells/mm$^3$ in CD4 cell counts have benefit from the combination treatment. 
Figure 10 shows the average treatment effect (ATE) and the $95$ \% Confidence Interval (CI) of the rules selected in Table 3. 
The $95$ \% CI for the ATE of any selected $8$  rules did not include $0$, indicating that the treatment group performed better for all selected rules. This result was consistent with that of the original clinical study [Hammer et al., 1996]. 
The point estimations of the ATE in all rules were higher than the overall ATE, and the confidence intervals of the rules \#$4, 5,$ and $6$ were higher than the overall ATE. It showed that these subgroups could benefit from the targeted treatment.  

\begin{figure}[htbp]
\begin{center}
\includegraphics[scale=0.42]{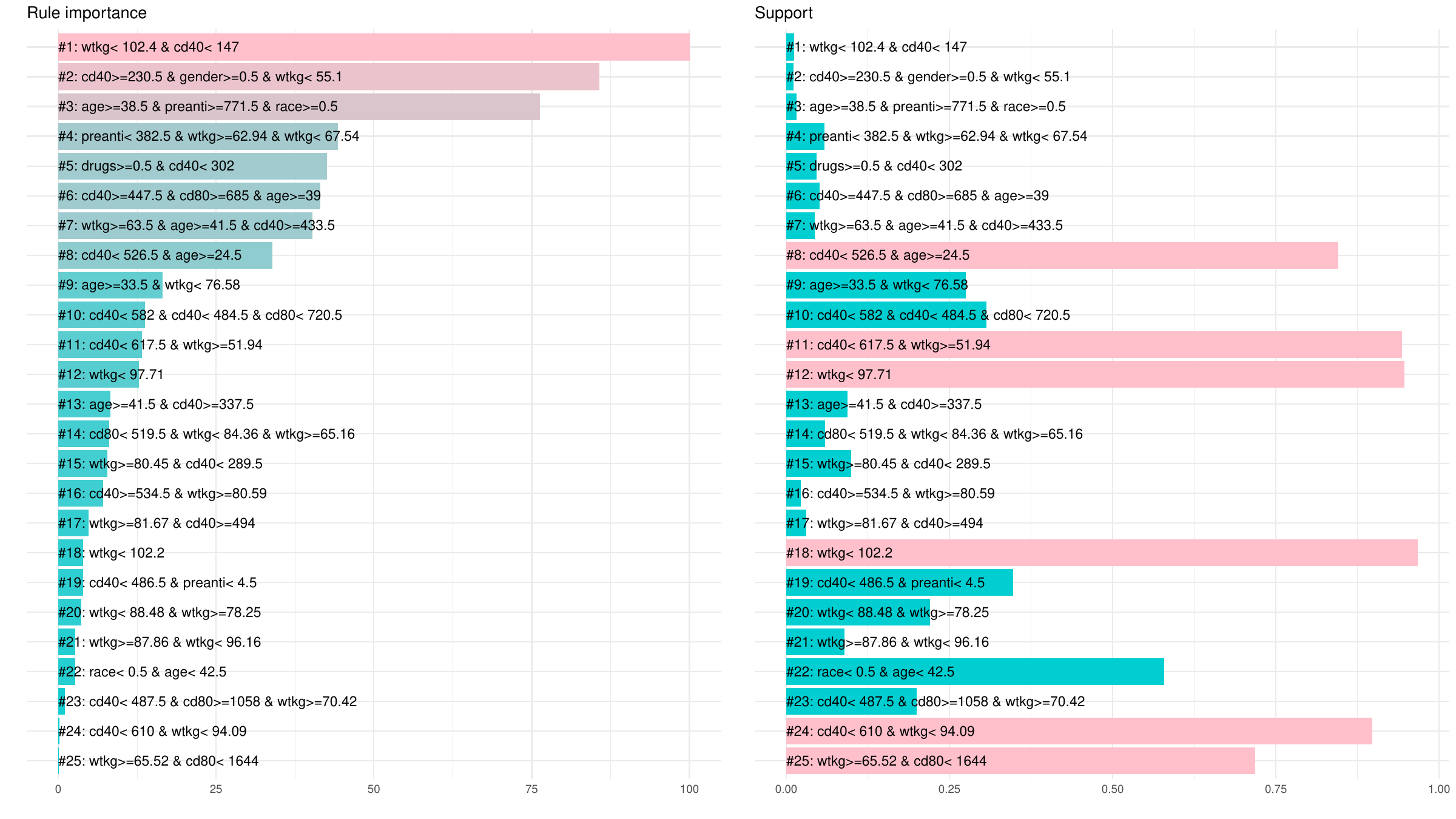}

\caption{Plots of rule importance and the support of the rule importance. The left bar plot describes the rule importance; the horizontal axis is the importance of the rules, and the vertical axis is the rules. The right bar plot is the support of the rule importance. The horizontal axis describes the support value which depict pink color if the value is more than $0.7$.}
\label{ruleimp}
\end{center}
\end{figure}

\begin{table}[H]
\begin{center}
\caption{Rule importance, HTE and its support of the proposed method.}\label{sup}
\scalebox{0.8}{
\begin{tabular}{rrrrrr} 
\hline
Rule \# & Rule & Rule Importance & HTE & Support   \\\hline
$1$ & wtkg $< 102.4$ \& cd40  $<147$ & $100.00$ & $71.44$ & $0.01$ & \\
$2$ & cd40 $\geq230.5$ \& gender $\geq 0.5$ \& wtkg $<55.1$ & $85.69$ & $61.21$ & $0.01$  \\
$3$ & age $\geq 38.5$ \& preanti $\geq 771.5$ \& race $\geq 0.5$  & $76.31$ & $54.51$ & $0.02$  \\
$4$ & preanti $< 382.5$ \& wtkg $\geq 62.94$ \& wtkg $<67.54$ & $44.33$ & $31.67$ & $0.06$ \\
$5$ & drugs $\geq0.5$ \& cd40 $< 302$  & $42.57$ & $30.41$ & $0.05$ \\
$6$ & cd40 $\geq447.5$ \& cd80 $\geq685$ \& age $\geq39$ & $41.54$ & $29.67$ & $0.05$ \\
$7$ & wtkg $\geq63.5$ \& age $\geq41.5$ \& cd40 $\geq433.5$ & $40.21$ & $28.72$ & $0.04$ \\
$8$ & cd40 $<5226.5$ \& age $\geq24.5$  & $33.88$ & $24.20$ & $0.85$ \\\hline
\end{tabular}
}
\end{center}
$\quad \quad \quad \quad \quad \quad \quad$ 
\end{table}

\begin{figure}[htbp]
\begin{center}
\includegraphics[scale=0.7]{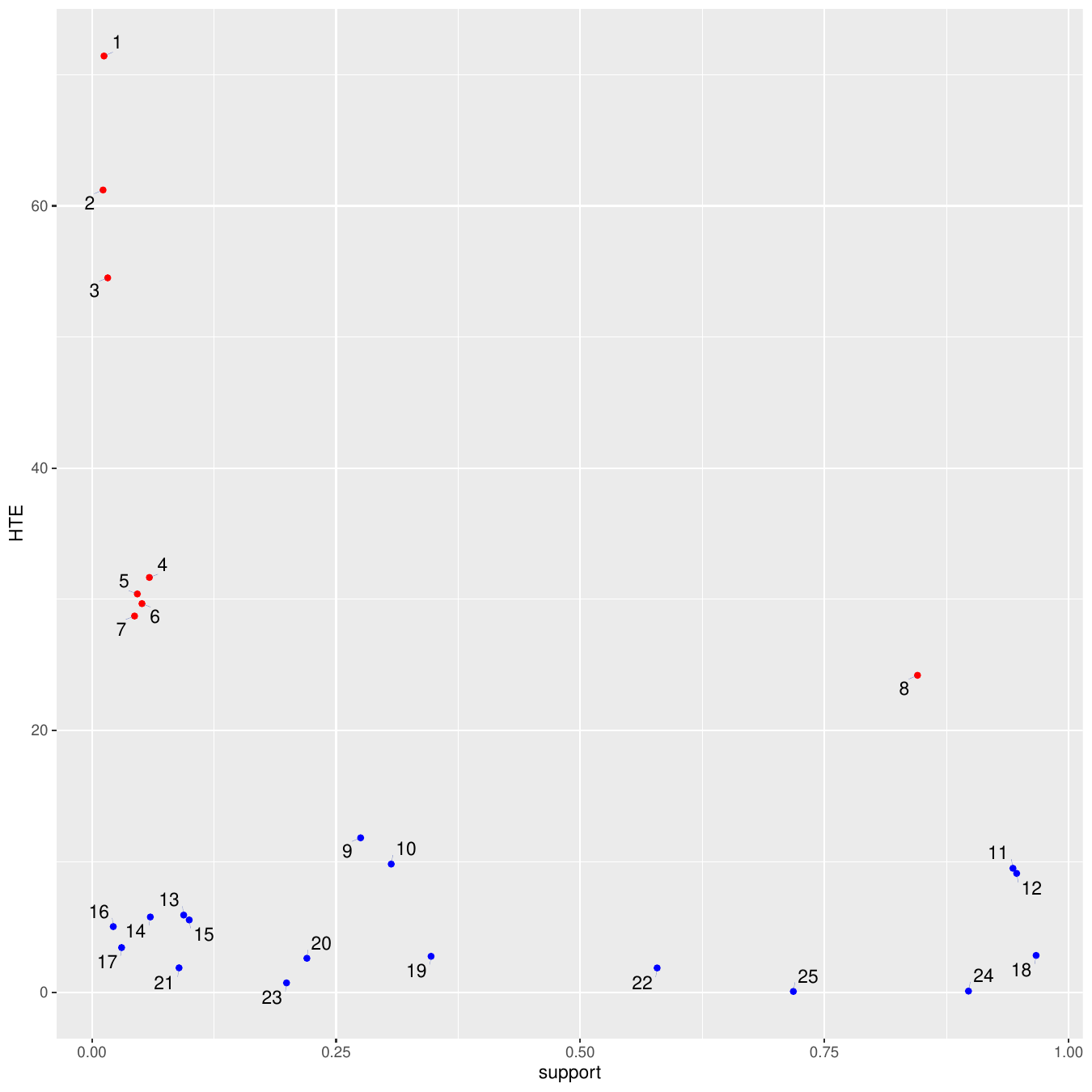}
\caption{Plot of the distribution of estimated HTE of selected $25$ rules and its support values. The vertical axis is HTE value and the horizontal axis is its support value. The number represents rules. The red point describes rules whose rule importance is more than its mean. }
\label{supdist}
\end{center}
\end{figure}

\begin{figure}[htbp]
\begin{center}
\includegraphics[scale=0.47]{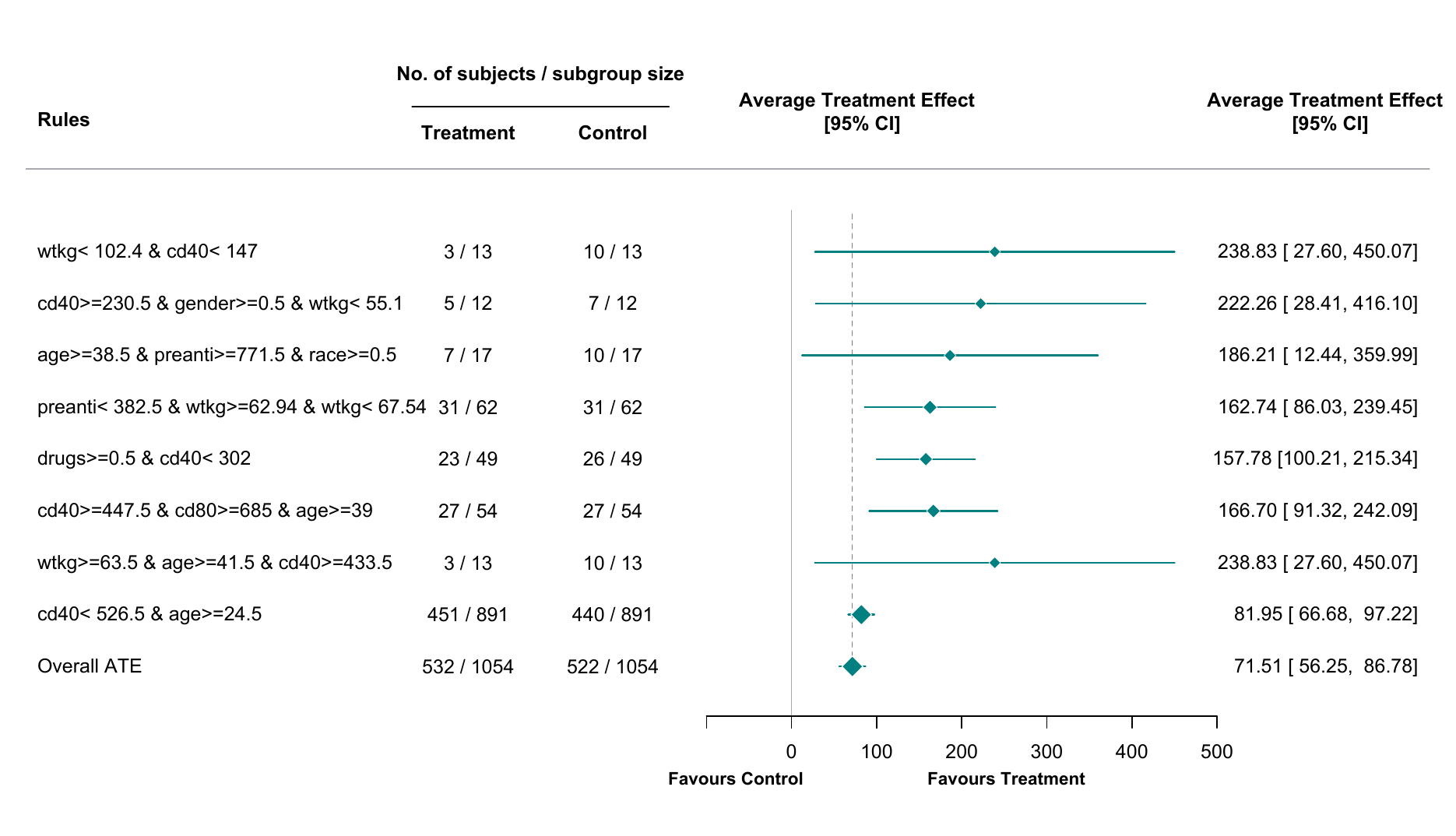}
\caption{Forest plot of the rules in Table 3. The rules selected in Table 3 are in the first column, the subgroup size of each treatment group is in the second and third columns,  the average treatment effect in each rules is depicted in the forth column and its $95$ \% confidence interval is in the fifth column. }
\label{forest}
\end{center}
\end{figure}

\newpage

\section{Discussion and conclusion}
This study proposed a novel framework based on RuleFit method to estimate the HTE.
The proposed framework adopted an S-learner to estimate the HTE by considering the main effect, which leads the interpretability the HTE with the form of rule.  
Through numerical simulation, we found that the proposed method estimated with the stability regardless of the number of covariates. 
Based on the Spearman's correlation coefficient results, we confirmed that the proposed method could capture the appropriate order of  magnitude of the correlation coefficients between the true treatment effects and predicted treatment effects. 
According to the scenarios, MSE of the proposed method showed better results than those of the compared methods in most scenarios, where the true treatment effects were formed of the threshold function. The relative bias in these scenarios were close to $0$ and it had high correlation. 
In scenarios where the treatment effects consisted of a combination of $\sin$ function and exponential function, the MSE values of the proposed method were close to those of Causal MARS and PTO forest. The proposed method maintained a low bias, and its correlation values were almost the same as those of Causal MARS and PTO forest in these situations. 
Conversely, when the true treatment effect comprised both linear and quadratic functions, MSE of Causal MARS was better than that of the proposed method.
Nonetheless, the proposed method exhibited superior performance over Causal MARS in scenarios where the true treatment effects included threshold functions or no treatment effects.

Therefore, it was found that the proposed method had a stable performance in several influences of the covariates on the outcome. 
Through its application to real clinical trial data, we confirmed the usefulness of the proposed method in terms of the interpretability of the estimated results using the estimated rules. 


\bibliography{references}  
\section*{Reference}

P. W. Holland. Statistics and causal inference. \textit{Journal of the American Statistical Association}, 81(396):945--960, 1986.\\
M. Gail and R. Simon. Testing for qualitative interactions between treatment effects and
patient subsets. \textit{Biometrics}, 41(2):361-372, 1985.\\
S. Wager and S. Athey. Estimation and inference of heterogeneous treatment effects using 
random forests. \textit{Journal of the American Statistical Association},113(523):1228--1242, 2018.\\
L. Breiman, R. Friedman, J. Olshen, and C. Stone. Classification and regression trees. \textit{Wardsworth}, 1984.\\
X. Su, C.L. Tsai, H. Wang, D.M. Nickerson, and B. Li. Subgroup analysis via recursive partitioning. \textit{Journal of Machine Learning Research}, 10(5):141--158, 2009.\\
S. Athey and G. Imbens. Recursive partitioning for heterogeneous causal effects. \textit{Proceedings of the National Academy of Sciences}, 113(27):7353--7360, 2016.\\
L. Breiman. Random forests. \textit{Machine Learning}, 45(1):5--32, 2001.\\
S. Athey, J. Tibshirani, and S. Wager. Generalized random forests. \textit{The Annals of Statistics}, 47(2):1148--1178, 2019.\\
S. Powers, J. Qian, K. Jung, A. Schuler, N.H. Shah, T. Hastie, and R. Tibshirani. Some methods for heterogeneous treatment effect estimation in high dimensions. \textit{Statistics in Medicine}, 37(11):1767--1787, 2018.\\
H.A. Chipman, E.I. George, and R.E. McCulloch. Bart: Bayesian additive regression trees. \textit{Annals of Applied Statistics}, 4(1):266--298, 2010.\\
J.L. Hill. Bayesian nonparametric modeling for causal inference. \textit{Journal of Computational and Graphical Statistics}, 20(1):217--240, 2011.\\
P. R. Hahn, J.S. Murray, and C.M. Carvalho. Bayesian Regression Tree Models for Causal Inference:
Regularization, Confounding, and Heterogeneous Effects (with Discussion). \textit{Bayesian Analysis}, 15(3):965--1056, 2020.\\
J.H. Friedman and B.E. Popescu. Predictive learning via rule ensembles. \textit{Annals of Applied Statistics}, 2(3):916--954, 2008.\\
F.J. Bargagli-Stoffi, R. Cadei, K. Lee, and F. Dominici. Causal rule ensemble: Interpretable discovery and inference of heterogeneous treatment effects. \textit{arXiv:2009.09036v4}, 2023.\\
S.R. Künzel, J.S. Sekhon, P.J. Bickel, and B. Yu. Metalearners for estimating heterogeneous treatment effects using machine learning. \textit{Proceedings of the national academy of sciences}, 116(19):4156--4165, 2019.\\
X. Nie and S. Wager. Quasi-oracle estimation of heterogeneous treatment effects. \textit{Biometrika}, 108(2):299-319, 2020.\\
M. Yuan and Y. Lin. Model selection and estimation in regression with grouped variables. \textit{Journal of the Royal Statistical Society: Series B (Statistical
Methodology)}, 68(1):49--67, 2006.\\
R. Tibshirani. Regression shrinkage and selection via the lasso. \textit{Journal of the Royal Statistical Society: Series B (Methodological)}, 58(1):267--288, 1996.\\
K. Wan, K. Tanioka, and T. Shimokawa. Rule ensemble method with adaptive group lasso for heterogeneous
treatment effect estimation. \textit{Statistics in Medicine}, online, 2023.\\
M. Fokkema. Fitting prediction rule ensembles with {R} package {pre}. \textit{Journal of Statistical Software}, 92(12):1--30, 2020.\\
J.H. Friedman. Greedy function approximation: a gradient boosting machine. \textit{Annals of Statistics}, 29(5):1189--1232, 2001.\\
T. Therneau and B. Atkinson. rpart: Recursive artitioning and Regression Trees, 2022. URL \url{https://CRAN.R-project.org/package=rpart}. R package version 4.1.16.\\
L. Tian, A.A. Alizadeh, A.J. Gentles, and R. Tibshirani. A simple method for estimating interactions between a treatment and a large number of covariates. \textit{Journal of the American Statistical Association}, 109(508):1575--1532, 2014.\\
J. Tibshirani, S. Athey, E. Sverdrup, and S. Wager. grf: Generalized Random Forests, 2022.  URL \url{https://CRAN.R-project.org/package=grf}. R package version 2.2.1.\\
S. Powers, J. Qian, T. Hastie, and R. Tibshirani. causalLearning: Methods for heterogeneous treatment effect
estimation, 2022. R package version 1.0.0. \\
S.M. Hammer, D.A. Katzenstein, M.D. Hughes, H. Gundacker, R.T. Schooley, R.H. Haubrich, W.K. Henry, M.M. Lederman, J.P. Phair, M. Niu, et al. A trial comparing nucleoside monotherapy with combination therapy in hiv-infected adults with cd4 cell counts from 200 to 500 per cubic
millimeter. \textit{New England Journal of Medicine}, 335(15):1081--1090, 1996.\\
M. Juraska, P.B. Gilbert, X. Lu, M. Zhang, M. Davidian, and A.A. Tsiatis. speff2trial: Semiparametric efficient estimation for a two-sample treatment effect, 2022. URL \url{https://CRAN.R-project.org/package=speff2trial}. R package version 1.0.5.\\
L.D. Saravolatz, D.L. Winslow, G. Collins, J.S. Hodges, C. Pettinelli, D.S. Stein, N. Markowitz, R. Reves, M.O. Loveless, L. Crane, M. Thompson, and D. Abrams. Zidovudine alone or in combination with didanosine or zalcitabine in hiv-infected patients with the acquired immunodeficiency syndrome or fewer than 200 cd4 cells per cubic millimeter. investigators for the terry beirn
community programs for clinical research on aids. \textit{The New England journal of medicine}, 335(15):1099--1106, 1996.\\

\end{document}